\documentclass[useAMS,usenatbib]{mn2e}

\title[Galaxy Zoo: the properties of merging galaxies (II)]{Galaxy Zoo: the properties of merging galaxies in the nearby Universe - local environments, colours, masses, star-formation rates and AGN activity}

\author[Darg et al.]{D. W. Darg,$^1$\thanks{Email: ddarg@astro.ox.ac.uk} S. Kaviraj,$^{2,1}$\thanks{Email: skaviraj@astro.ox.ac.uk} C. J. Lintott,$^1$\thanks{Email: cjl@astro.ox.ac.uk} K. Schawinski,$^{3}$ M. Sarzi,$^4$ S. Bamford,$^5$ \newauthor J. Silk,$^1$ D. Andreescu,$^6$ P. Murray,$^{7}$ R. C. Nichol,$^8$ M. J. Raddick,$^{9}$ \newauthor A. Slosar,$^{10}$ A. S. Szalay,$^{9}$ D. Thomas,$^8$ J. Vandenberg.$^{9}$ \thanks{This publication has been made possible by the participation of more than 140,000 volunteers in the Galaxy Zoo project. Their contributions are individually acknowledged at \texttt{http://www.galaxyzoo.org/Volunteers.aspx}.}\\ \\
$^{1}$ Department of Physics, University of Oxford, Keble Road, Oxford, OX1 3RH, UK\\
$^{2}$ Mullard Space Science Laboratory, University College London, Holmbury St. Mary, Dorking, Surrey RH5 6NT, UK \\
$^{3}$ Yale Center for Astronomy and Astrophysics, Yale University, P.O. Box 208121, New Haven, CT 06520, USA\\
$^{4}$ Centre for Astrophysics Research, University of Hertfordshire, Hatfield, AL10 9AB, UK	\\
$^{5}$ Centre for Astronomy and Particle Theory, University of Nottingham, University Park, Nottingham, NG7 2RD, UK \\
$^{6}$ LinkLab, 4506 Graystone Ave., Bronx, NY 10471, USA\\
$^{7}$ Fingerprint Digital Media, 9 Victoria Close, Newtownards, Co. Down, Northern Ireland, BT23 7GY, UK\\
$^{8}$ Institute of Cosmology and Gravitation, University of Portsmouth, Mercantile House, Hampshire Terrace, Portsmouth, PO1 2EG, UK\\
$^{9}$ Department of Physics and Astronomy, Johns Hopkins University, 3400 N. Charles St., Baltimore, MD 21218, USA\\
$^{10}$ Berkeley Center for Cosmological Physics, Lawrence Berkeley National Laboratory and Physics Department, University of California, \\Berkeley CA 94720, USA\\
}

\usepackage{graphicx}   
\usepackage{wasysym}	
\usepackage{amsfonts}	%
\usepackage{color}
\definecolor{AliceBlue}{rgb}{0.94,0.97,1.00}
\definecolor{AntiqueWhite1}{rgb}{1.00,0.94,0.86}
\definecolor{AntiqueWhite2}{rgb}{0.93,0.87,0.80}
\definecolor{AntiqueWhite3}{rgb}{0.80,0.75,0.69}
\definecolor{AntiqueWhite4}{rgb}{0.55,0.51,0.47}
\definecolor{AntiqueWhite}{rgb}{0.98,0.92,0.84}
\definecolor{BlanchedAlmond}{rgb}{1.00,0.92,0.80}
\definecolor{BlueViolet}{rgb}{0.54,0.17,0.89}
\definecolor{CadetBlue1}{rgb}{0.60,0.96,1.00}
\definecolor{CadetBlue2}{rgb}{0.56,0.90,0.93}
\definecolor{CadetBlue3}{rgb}{0.48,0.77,0.80}
\definecolor{CadetBlue4}{rgb}{0.33,0.53,0.55}
\definecolor{CadetBlue}{rgb}{0.37,0.62,0.63}
\definecolor{CornflowerBlue}{rgb}{0.39,0.58,0.93}
\definecolor{DarkBlue}{rgb}{0.00,0.00,0.55}
\definecolor{DarkCyan}{rgb}{0.00,0.55,0.55}
\definecolor{DarkGoldenrod1}{rgb}{1.00,0.73,0.06}
\definecolor{DarkGoldenrod2}{rgb}{0.93,0.68,0.05}
\definecolor{DarkGoldenrod3}{rgb}{0.80,0.58,0.05}
\definecolor{DarkGoldenrod4}{rgb}{0.55,0.40,0.03}
\definecolor{DarkGoldenrod}{rgb}{0.72,0.53,0.04}
\definecolor{DarkGray}{rgb}{0.66,0.66,0.66}
\definecolor{DarkGreen}{rgb}{0.00,0.39,0.00}
\definecolor{DarkGrey}{rgb}{0.66,0.66,0.66}
\definecolor{DarkKhaki}{rgb}{0.74,0.72,0.42}
\definecolor{DarkMagenta}{rgb}{0.55,0.00,0.55}
\definecolor{DarkOliveGreen1}{rgb}{0.79,1.00,0.44}
\definecolor{DarkOliveGreen2}{rgb}{0.74,0.93,0.41}
\definecolor{DarkOliveGreen3}{rgb}{0.64,0.80,0.35}
\definecolor{DarkOliveGreen4}{rgb}{0.43,0.55,0.24}
\definecolor{DarkOliveGreen}{rgb}{0.33,0.42,0.18}
\definecolor{DarkOrange1}{rgb}{1.00,0.50,0.00}
\definecolor{DarkOrange2}{rgb}{0.93,0.46,0.00}
\definecolor{DarkOrange3}{rgb}{0.80,0.40,0.00}
\definecolor{DarkOrange4}{rgb}{0.55,0.27,0.00}
\definecolor{DarkOrange}{rgb}{1.00,0.55,0.00}
\definecolor{DarkOrchid1}{rgb}{0.75,0.24,1.00}
\definecolor{DarkOrchid2}{rgb}{0.70,0.23,0.93}
\definecolor{DarkOrchid3}{rgb}{0.60,0.20,0.80}
\definecolor{DarkOrchid4}{rgb}{0.41,0.13,0.55}
\definecolor{DarkOrchid}{rgb}{0.60,0.20,0.80}
\definecolor{DarkRed}{rgb}{0.55,0.00,0.00}
\definecolor{DarkSalmon}{rgb}{0.91,0.59,0.48}
\definecolor{DarkSeaGreen1}{rgb}{0.76,1.00,0.76}
\definecolor{DarkSeaGreen2}{rgb}{0.71,0.93,0.71}
\definecolor{DarkSeaGreen3}{rgb}{0.61,0.80,0.61}
\definecolor{DarkSeaGreen4}{rgb}{0.41,0.55,0.41}
\definecolor{DarkSeaGreen}{rgb}{0.56,0.74,0.56}
\definecolor{DarkSlateBlue}{rgb}{0.28,0.24,0.55}
\definecolor{DarkSlateGray1}{rgb}{0.59,1.00,1.00}
\definecolor{DarkSlateGray2}{rgb}{0.55,0.93,0.93}
\definecolor{DarkSlateGray3}{rgb}{0.47,0.80,0.80}
\definecolor{DarkSlateGray4}{rgb}{0.32,0.55,0.55}
\definecolor{DarkSlateGray}{rgb}{0.18,0.31,0.31}
\definecolor{DarkSlateGrey}{rgb}{0.18,0.31,0.31}
\definecolor{DarkTurquoise}{rgb}{0.00,0.81,0.82}
\definecolor{DarkViolet}{rgb}{0.58,0.00,0.83}
\definecolor{DeepPink1}{rgb}{1.00,0.08,0.58}
\definecolor{DeepPink2}{rgb}{0.93,0.07,0.54}
\definecolor{DeepPink3}{rgb}{0.80,0.06,0.46}
\definecolor{DeepPink4}{rgb}{0.55,0.04,0.31}
\definecolor{DeepPink}{rgb}{1.00,0.08,0.58}
\definecolor{DeepSkyBlue1}{rgb}{0.00,0.75,1.00}
\definecolor{DeepSkyBlue2}{rgb}{0.00,0.70,0.93}
\definecolor{DeepSkyBlue3}{rgb}{0.00,0.60,0.80}
\definecolor{DeepSkyBlue4}{rgb}{0.00,0.41,0.55}
\definecolor{DeepSkyBlue}{rgb}{0.00,0.75,1.00}
\definecolor{DimGray}{rgb}{0.41,0.41,0.41}
\definecolor{DimGrey}{rgb}{0.41,0.41,0.41}
\definecolor{DodgerBlue1}{rgb}{0.12,0.56,1.00}
\definecolor{DodgerBlue2}{rgb}{0.11,0.53,0.93}
\definecolor{DodgerBlue3}{rgb}{0.09,0.45,0.80}
\definecolor{DodgerBlue4}{rgb}{0.06,0.31,0.55}
\definecolor{DodgerBlue}{rgb}{0.12,0.56,1.00}
\definecolor{FloralWhite}{rgb}{1.00,0.98,0.94}
\definecolor{ForestGreen}{rgb}{0.13,0.55,0.13}
\definecolor{GhostWhite}{rgb}{0.97,0.97,1.00}
\definecolor{GreenYellow}{rgb}{0.68,1.00,0.18}
\definecolor{HotPink1}{rgb}{1.00,0.43,0.71}
\definecolor{HotPink2}{rgb}{0.93,0.42,0.65}
\definecolor{HotPink3}{rgb}{0.80,0.38,0.56}
\definecolor{HotPink4}{rgb}{0.55,0.23,0.38}
\definecolor{HotPink}{rgb}{1.00,0.41,0.71}
\definecolor{IndianRed1}{rgb}{1.00,0.42,0.42}
\definecolor{IndianRed2}{rgb}{0.93,0.39,0.39}
\definecolor{IndianRed3}{rgb}{0.80,0.33,0.33}
\definecolor{IndianRed4}{rgb}{0.55,0.23,0.23}
\definecolor{IndianRed}{rgb}{0.80,0.36,0.36}
\definecolor{LavenderBlush1}{rgb}{1.00,0.94,0.96}
\definecolor{LavenderBlush2}{rgb}{0.93,0.88,0.90}
\definecolor{LavenderBlush3}{rgb}{0.80,0.76,0.77}
\definecolor{LavenderBlush4}{rgb}{0.55,0.51,0.53}
\definecolor{LavenderBlush}{rgb}{1.00,0.94,0.96}
\definecolor{LawnGreen}{rgb}{0.49,0.99,0.00}
\definecolor{LemonChiffon1}{rgb}{1.00,0.98,0.80}
\definecolor{LemonChiffon2}{rgb}{0.93,0.91,0.75}
\definecolor{LemonChiffon3}{rgb}{0.80,0.79,0.65}
\definecolor{LemonChiffon4}{rgb}{0.55,0.54,0.44}
\definecolor{LemonChiffon}{rgb}{1.00,0.98,0.80}
\definecolor{LightBlue1}{rgb}{0.75,0.94,1.00}
\definecolor{LightBlue2}{rgb}{0.70,0.87,0.93}
\definecolor{LightBlue3}{rgb}{0.60,0.75,0.80}
\definecolor{LightBlue4}{rgb}{0.41,0.51,0.55}
\definecolor{LightBlue}{rgb}{0.68,0.85,0.90}
\definecolor{LightCoral}{rgb}{0.94,0.50,0.50}
\definecolor{LightCyan1}{rgb}{0.88,1.00,1.00}
\definecolor{LightCyan2}{rgb}{0.82,0.93,0.93}
\definecolor{LightCyan3}{rgb}{0.71,0.80,0.80}
\definecolor{LightCyan4}{rgb}{0.48,0.55,0.55}
\definecolor{LightCyan}{rgb}{0.88,1.00,1.00}
\definecolor{LightGoldenrod1}{rgb}{1.00,0.93,0.55}
\definecolor{LightGoldenrod2}{rgb}{0.93,0.86,0.51}
\definecolor{LightGoldenrod3}{rgb}{0.80,0.75,0.44}
\definecolor{LightGoldenrod4}{rgb}{0.55,0.51,0.30}
\definecolor{LightGoldenrodYellow}{rgb}{0.98,0.98,0.82}
\definecolor{LightGoldenrod}{rgb}{0.93,0.87,0.51}
\definecolor{LightGray}{rgb}{0.83,0.83,0.83}
\definecolor{LightGreen}{rgb}{0.56,0.93,0.56}
\definecolor{LightGrey}{rgb}{0.83,0.83,0.83}
\definecolor{LightPink1}{rgb}{1.00,0.68,0.73}
\definecolor{LightPink2}{rgb}{0.93,0.64,0.68}
\definecolor{LightPink3}{rgb}{0.80,0.55,0.58}
\definecolor{LightPink4}{rgb}{0.55,0.37,0.40}
\definecolor{LightPink}{rgb}{1.00,0.71,0.76}
\definecolor{LightSalmon1}{rgb}{1.00,0.63,0.48}
\definecolor{LightSalmon2}{rgb}{0.93,0.58,0.45}
\definecolor{LightSalmon3}{rgb}{0.80,0.51,0.38}
\definecolor{LightSalmon4}{rgb}{0.55,0.34,0.26}
\definecolor{LightSalmon}{rgb}{1.00,0.63,0.48}
\definecolor{LightSeaGreen}{rgb}{0.13,0.70,0.67}
\definecolor{LightSkyBlue1}{rgb}{0.69,0.89,1.00}
\definecolor{LightSkyBlue2}{rgb}{0.64,0.83,0.93}
\definecolor{LightSkyBlue3}{rgb}{0.55,0.71,0.80}
\definecolor{LightSkyBlue4}{rgb}{0.38,0.48,0.55}
\definecolor{LightSkyBlue}{rgb}{0.53,0.81,0.98}
\definecolor{LightSlateBlue}{rgb}{0.52,0.44,1.00}
\definecolor{LightSlateGray}{rgb}{0.47,0.53,0.60}
\definecolor{LightSlateGrey}{rgb}{0.47,0.53,0.60}
\definecolor{LightSteelBlue1}{rgb}{0.79,0.88,1.00}
\definecolor{LightSteelBlue2}{rgb}{0.74,0.82,0.93}
\definecolor{LightSteelBlue3}{rgb}{0.64,0.71,0.80}
\definecolor{LightSteelBlue4}{rgb}{0.43,0.48,0.55}
\definecolor{LightSteelBlue}{rgb}{0.69,0.77,0.87}
\definecolor{LightYellow1}{rgb}{1.00,1.00,0.88}
\definecolor{LightYellow2}{rgb}{0.93,0.93,0.82}
\definecolor{LightYellow3}{rgb}{0.80,0.80,0.71}
\definecolor{LightYellow4}{rgb}{0.55,0.55,0.48}
\definecolor{LightYellow}{rgb}{1.00,1.00,0.88}
\definecolor{LimeGreen}{rgb}{0.20,0.80,0.20}
\definecolor{MediumAquamarine}{rgb}{0.40,0.80,0.67}
\definecolor{MediumBlue}{rgb}{0.00,0.00,0.80}
\definecolor{MediumOrchid1}{rgb}{0.88,0.40,1.00}
\definecolor{MediumOrchid2}{rgb}{0.82,0.37,0.93}
\definecolor{MediumOrchid3}{rgb}{0.71,0.32,0.80}
\definecolor{MediumOrchid4}{rgb}{0.48,0.22,0.55}
\definecolor{MediumOrchid}{rgb}{0.73,0.33,0.83}
\definecolor{MediumPurple1}{rgb}{0.67,0.51,1.00}
\definecolor{MediumPurple2}{rgb}{0.62,0.47,0.93}
\definecolor{MediumPurple3}{rgb}{0.54,0.41,0.80}
\definecolor{MediumPurple4}{rgb}{0.36,0.28,0.55}
\definecolor{MediumPurple}{rgb}{0.58,0.44,0.86}
\definecolor{MediumSeaGreen}{rgb}{0.24,0.70,0.44}
\definecolor{MediumSlateBlue}{rgb}{0.48,0.41,0.93}
\definecolor{MediumSpringGreen}{rgb}{0.00,0.98,0.60}
\definecolor{MediumTurquoise}{rgb}{0.28,0.82,0.80}
\definecolor{MediumVioletRed}{rgb}{0.78,0.08,0.52}
\definecolor{MidnightBlue}{rgb}{0.10,0.10,0.44}
\definecolor{MintCream}{rgb}{0.96,1.00,0.98}
\definecolor{MistyRose1}{rgb}{1.00,0.89,0.88}
\definecolor{MistyRose2}{rgb}{0.93,0.84,0.82}
\definecolor{MistyRose3}{rgb}{0.80,0.72,0.71}
\definecolor{MistyRose4}{rgb}{0.55,0.49,0.48}
\definecolor{MistyRose}{rgb}{1.00,0.89,0.88}
\definecolor{NavajoWhite1}{rgb}{1.00,0.87,0.68}
\definecolor{NavajoWhite2}{rgb}{0.93,0.81,0.63}
\definecolor{NavajoWhite3}{rgb}{0.80,0.70,0.55}
\definecolor{NavajoWhite4}{rgb}{0.55,0.47,0.37}
\definecolor{NavajoWhite}{rgb}{1.00,0.87,0.68}
\definecolor{NavyBlue}{rgb}{0.00,0.00,0.50}
\definecolor{OldLace}{rgb}{0.99,0.96,0.90}
\definecolor{OliveDrab1}{rgb}{0.75,1.00,0.24}
\definecolor{OliveDrab2}{rgb}{0.70,0.93,0.23}
\definecolor{OliveDrab3}{rgb}{0.60,0.80,0.20}
\definecolor{OliveDrab4}{rgb}{0.41,0.55,0.13}
\definecolor{OliveDrab}{rgb}{0.42,0.56,0.14}
\definecolor{OrangeRed1}{rgb}{1.00,0.27,0.00}
\definecolor{OrangeRed2}{rgb}{0.93,0.25,0.00}
\definecolor{OrangeRed3}{rgb}{0.80,0.22,0.00}
\definecolor{OrangeRed4}{rgb}{0.55,0.15,0.00}
\definecolor{OrangeRed}{rgb}{1.00,0.27,0.00}
\definecolor{PaleGoldenrod}{rgb}{0.93,0.91,0.67}
\definecolor{PaleGreen1}{rgb}{0.60,1.00,0.60}
\definecolor{PaleGreen2}{rgb}{0.56,0.93,0.56}
\definecolor{PaleGreen3}{rgb}{0.49,0.80,0.49}
\definecolor{PaleGreen4}{rgb}{0.33,0.55,0.33}
\definecolor{PaleGreen}{rgb}{0.60,0.98,0.60}
\definecolor{PaleTurquoise1}{rgb}{0.73,1.00,1.00}
\definecolor{PaleTurquoise2}{rgb}{0.68,0.93,0.93}
\definecolor{PaleTurquoise3}{rgb}{0.59,0.80,0.80}
\definecolor{PaleTurquoise4}{rgb}{0.40,0.55,0.55}
\definecolor{PaleTurquoise}{rgb}{0.69,0.93,0.93}
\definecolor{PaleVioletRed1}{rgb}{1.00,0.51,0.67}
\definecolor{PaleVioletRed2}{rgb}{0.93,0.47,0.62}
\definecolor{PaleVioletRed3}{rgb}{0.80,0.41,0.54}
\definecolor{PaleVioletRed4}{rgb}{0.55,0.28,0.36}
\definecolor{PaleVioletRed}{rgb}{0.86,0.44,0.58}
\definecolor{PapayaWhip}{rgb}{1.00,0.94,0.84}
\definecolor{PeachPuff1}{rgb}{1.00,0.85,0.73}
\definecolor{PeachPuff2}{rgb}{0.93,0.80,0.68}
\definecolor{PeachPuff3}{rgb}{0.80,0.69,0.58}
\definecolor{PeachPuff4}{rgb}{0.55,0.47,0.40}
\definecolor{PeachPuff}{rgb}{1.00,0.85,0.73}
\definecolor{PowderBlue}{rgb}{0.69,0.88,0.90}
\definecolor{RosyBrown1}{rgb}{1.00,0.76,0.76}
\definecolor{RosyBrown2}{rgb}{0.93,0.71,0.71}
\definecolor{RosyBrown3}{rgb}{0.80,0.61,0.61}
\definecolor{RosyBrown4}{rgb}{0.55,0.41,0.41}
\definecolor{RosyBrown}{rgb}{0.74,0.56,0.56}
\definecolor{RoyalBlue1}{rgb}{0.28,0.46,1.00}
\definecolor{RoyalBlue2}{rgb}{0.26,0.43,0.93}
\definecolor{RoyalBlue3}{rgb}{0.23,0.37,0.80}
\definecolor{RoyalBlue4}{rgb}{0.15,0.25,0.55}
\definecolor{RoyalBlue}{rgb}{0.25,0.41,0.88}
\definecolor{SaddleBrown}{rgb}{0.55,0.27,0.07}
\definecolor{SandyBrown}{rgb}{0.96,0.64,0.38}
\definecolor{SeaGreen1}{rgb}{0.33,1.00,0.62}
\definecolor{SeaGreen2}{rgb}{0.31,0.93,0.58}
\definecolor{SeaGreen3}{rgb}{0.26,0.80,0.50}
\definecolor{SeaGreen4}{rgb}{0.18,0.55,0.34}
\definecolor{SeaGreen}{rgb}{0.18,0.55,0.34}
\definecolor{SkyBlue1}{rgb}{0.53,0.81,1.00}
\definecolor{SkyBlue2}{rgb}{0.49,0.75,0.93}
\definecolor{SkyBlue3}{rgb}{0.42,0.65,0.80}
\definecolor{SkyBlue4}{rgb}{0.29,0.44,0.55}
\definecolor{SkyBlue}{rgb}{0.53,0.81,0.92}
\definecolor{SlateBlue1}{rgb}{0.51,0.44,1.00}
\definecolor{SlateBlue2}{rgb}{0.48,0.40,0.93}
\definecolor{SlateBlue3}{rgb}{0.41,0.35,0.80}
\definecolor{SlateBlue4}{rgb}{0.28,0.24,0.55}
\definecolor{SlateBlue}{rgb}{0.42,0.35,0.80}
\definecolor{SlateGray1}{rgb}{0.78,0.89,1.00}
\definecolor{SlateGray2}{rgb}{0.73,0.83,0.93}
\definecolor{SlateGray3}{rgb}{0.62,0.71,0.80}
\definecolor{SlateGray4}{rgb}{0.42,0.48,0.55}
\definecolor{SlateGray}{rgb}{0.44,0.50,0.56}
\definecolor{SlateGrey}{rgb}{0.44,0.50,0.56}
\definecolor{SpringGreen1}{rgb}{0.00,1.00,0.50}
\definecolor{SpringGreen2}{rgb}{0.00,0.93,0.46}
\definecolor{SpringGreen3}{rgb}{0.00,0.80,0.40}
\definecolor{SpringGreen4}{rgb}{0.00,0.55,0.27}
\definecolor{SpringGreen}{rgb}{0.00,1.00,0.50}
\definecolor{SteelBlue1}{rgb}{0.39,0.72,1.00}
\definecolor{SteelBlue2}{rgb}{0.36,0.67,0.93}
\definecolor{SteelBlue3}{rgb}{0.31,0.58,0.80}
\definecolor{SteelBlue4}{rgb}{0.21,0.39,0.55}
\definecolor{SteelBlue}{rgb}{0.27,0.51,0.71}
\definecolor{VioletRed1}{rgb}{1.00,0.24,0.59}
\definecolor{VioletRed2}{rgb}{0.93,0.23,0.55}
\definecolor{VioletRed3}{rgb}{0.80,0.20,0.47}
\definecolor{VioletRed4}{rgb}{0.55,0.13,0.32}
\definecolor{VioletRed}{rgb}{0.82,0.13,0.56}
\definecolor{WhiteSmoke}{rgb}{0.96,0.96,0.96}
\definecolor{YellowGreen}{rgb}{0.60,0.80,0.20}
\definecolor{aliceblue}{rgb}{0.94,0.97,1.00}
\definecolor{antiquewhite}{rgb}{0.98,0.92,0.84}
\definecolor{aquamarine1}{rgb}{0.50,1.00,0.83}
\definecolor{aquamarine2}{rgb}{0.46,0.93,0.78}
\definecolor{aquamarine3}{rgb}{0.40,0.80,0.67}
\definecolor{aquamarine4}{rgb}{0.27,0.55,0.45}
\definecolor{aquamarine}{rgb}{0.50,1.00,0.83}
\definecolor{azure1}{rgb}{0.94,1.00,1.00}
\definecolor{azure2}{rgb}{0.88,0.93,0.93}
\definecolor{azure3}{rgb}{0.76,0.80,0.80}
\definecolor{azure4}{rgb}{0.51,0.55,0.55}
\definecolor{azure}{rgb}{0.94,1.00,1.00}
\definecolor{beige}{rgb}{0.96,0.96,0.86}
\definecolor{bisque1}{rgb}{1.00,0.89,0.77}
\definecolor{bisque2}{rgb}{0.93,0.84,0.72}
\definecolor{bisque3}{rgb}{0.80,0.72,0.62}
\definecolor{bisque4}{rgb}{0.55,0.49,0.42}
\definecolor{bisque}{rgb}{1.00,0.89,0.77}
\definecolor{black}{rgb}{0.00,0.00,0.00}
\definecolor{blanchedalmond}{rgb}{1.00,0.92,0.80}
\definecolor{blue1}{rgb}{0.00,0.00,1.00}
\definecolor{blue2}{rgb}{0.00,0.00,0.93}
\definecolor{blue3}{rgb}{0.00,0.00,0.80}
\definecolor{blue4}{rgb}{0.00,0.00,0.55}
\definecolor{blueviolet}{rgb}{0.54,0.17,0.89}
\definecolor{blue}{rgb}{0.00,0.00,1.00}
\definecolor{brown1}{rgb}{1.00,0.25,0.25}
\definecolor{brown2}{rgb}{0.93,0.23,0.23}
\definecolor{brown3}{rgb}{0.80,0.20,0.20}
\definecolor{brown4}{rgb}{0.55,0.14,0.14}
\definecolor{brown}{rgb}{0.65,0.16,0.16}
\definecolor{burlywood1}{rgb}{1.00,0.83,0.61}
\definecolor{burlywood2}{rgb}{0.93,0.77,0.57}
\definecolor{burlywood3}{rgb}{0.80,0.67,0.49}
\definecolor{burlywood4}{rgb}{0.55,0.45,0.33}
\definecolor{burlywood}{rgb}{0.87,0.72,0.53}
\definecolor{cadetblue}{rgb}{0.37,0.62,0.63}
\definecolor{chartreuse1}{rgb}{0.50,1.00,0.00}
\definecolor{chartreuse2}{rgb}{0.46,0.93,0.00}
\definecolor{chartreuse3}{rgb}{0.40,0.80,0.00}
\definecolor{chartreuse4}{rgb}{0.27,0.55,0.00}
\definecolor{chartreuse}{rgb}{0.50,1.00,0.00}
\definecolor{chocolate1}{rgb}{1.00,0.50,0.14}
\definecolor{chocolate2}{rgb}{0.93,0.46,0.13}
\definecolor{chocolate3}{rgb}{0.80,0.40,0.11}
\definecolor{chocolate4}{rgb}{0.55,0.27,0.07}
\definecolor{chocolate}{rgb}{0.82,0.41,0.12}
\definecolor{coral1}{rgb}{1.00,0.45,0.34}
\definecolor{coral2}{rgb}{0.93,0.42,0.31}
\definecolor{coral3}{rgb}{0.80,0.36,0.27}
\definecolor{coral4}{rgb}{0.55,0.24,0.18}
\definecolor{coral}{rgb}{1.00,0.50,0.31}
\definecolor{cornflowerblue}{rgb}{0.39,0.58,0.93}
\definecolor{cornsilk1}{rgb}{1.00,0.97,0.86}
\definecolor{cornsilk2}{rgb}{0.93,0.91,0.80}
\definecolor{cornsilk3}{rgb}{0.80,0.78,0.69}
\definecolor{cornsilk4}{rgb}{0.55,0.53,0.47}
\definecolor{cornsilk}{rgb}{1.00,0.97,0.86}
\definecolor{cyan1}{rgb}{0.00,1.00,1.00}
\definecolor{cyan2}{rgb}{0.00,0.93,0.93}
\definecolor{cyan3}{rgb}{0.00,0.80,0.80}
\definecolor{cyan4}{rgb}{0.00,0.55,0.55}
\definecolor{cyan}{rgb}{0.00,1.00,1.00}
\definecolor{darkblue}{rgb}{0.00,0.00,0.55}
\definecolor{darkcyan}{rgb}{0.00,0.55,0.55}
\definecolor{darkgoldenrod}{rgb}{0.72,0.53,0.04}
\definecolor{darkgray}{rgb}{0.66,0.66,0.66}
\definecolor{darkgreen}{rgb}{0.00,0.39,0.00}
\definecolor{darkgrey}{rgb}{0.66,0.66,0.66}
\definecolor{darkkhaki}{rgb}{0.74,0.72,0.42}
\definecolor{darkmagenta}{rgb}{0.55,0.00,0.55}
\definecolor{darkolive}{rgb}{0.33,0.42,0.18}
\definecolor{darkorange}{rgb}{1.00,0.55,0.00}
\definecolor{darkorchid}{rgb}{0.60,0.20,0.80}
\definecolor{darkred}{rgb}{0.55,0.00,0.00}
\definecolor{darksalmon}{rgb}{0.91,0.59,0.48}
\definecolor{darksea}{rgb}{0.56,0.74,0.56}
\definecolor{darkslate}{rgb}{0.18,0.31,0.31}
\definecolor{darkslate}{rgb}{0.18,0.31,0.31}
\definecolor{darkslate}{rgb}{0.28,0.24,0.55}
\definecolor{darkturquoise}{rgb}{0.00,0.81,0.82}
\definecolor{darkviolet}{rgb}{0.58,0.00,0.83}
\definecolor{deeppink}{rgb}{1.00,0.08,0.58}
\definecolor{deepsky}{rgb}{0.00,0.75,1.00}
\definecolor{dimgray}{rgb}{0.41,0.41,0.41}
\definecolor{dimgrey}{rgb}{0.41,0.41,0.41}
\definecolor{dodgerblue}{rgb}{0.12,0.56,1.00}
\definecolor{firebrick1}{rgb}{1.00,0.19,0.19}
\definecolor{firebrick2}{rgb}{0.93,0.17,0.17}
\definecolor{firebrick3}{rgb}{0.80,0.15,0.15}
\definecolor{firebrick4}{rgb}{0.55,0.10,0.10}
\definecolor{firebrick}{rgb}{0.70,0.13,0.13}
\definecolor{floralwhite}{rgb}{1.00,0.98,0.94}
\definecolor{forestgreen}{rgb}{0.13,0.55,0.13}
\definecolor{gainsboro}{rgb}{0.86,0.86,0.86}
\definecolor{ghostwhite}{rgb}{0.97,0.97,1.00}
\definecolor{gold1}{rgb}{1.00,0.84,0.00}
\definecolor{gold2}{rgb}{0.93,0.79,0.00}
\definecolor{gold3}{rgb}{0.80,0.68,0.00}
\definecolor{gold4}{rgb}{0.55,0.46,0.00}
\definecolor{goldenrod1}{rgb}{1.00,0.76,0.15}
\definecolor{goldenrod2}{rgb}{0.93,0.71,0.13}
\definecolor{goldenrod3}{rgb}{0.80,0.61,0.11}
\definecolor{goldenrod4}{rgb}{0.55,0.41,0.08}
\definecolor{goldenrod}{rgb}{0.85,0.65,0.13}
\definecolor{gold}{rgb}{1.00,0.84,0.00}
\definecolor{gray0}{rgb}{0.00,0.00,0.00}
\definecolor{gray100}{rgb}{1.00,1.00,1.00}
\definecolor{gray10}{rgb}{0.10,0.10,0.10}
\definecolor{gray11}{rgb}{0.11,0.11,0.11}
\definecolor{gray12}{rgb}{0.12,0.12,0.12}
\definecolor{gray13}{rgb}{0.13,0.13,0.13}
\definecolor{gray14}{rgb}{0.14,0.14,0.14}
\definecolor{gray15}{rgb}{0.15,0.15,0.15}
\definecolor{gray16}{rgb}{0.16,0.16,0.16}
\definecolor{gray17}{rgb}{0.17,0.17,0.17}
\definecolor{gray18}{rgb}{0.18,0.18,0.18}
\definecolor{gray19}{rgb}{0.19,0.19,0.19}
\definecolor{gray1}{rgb}{0.01,0.01,0.01}
\definecolor{gray20}{rgb}{0.20,0.20,0.20}
\definecolor{gray21}{rgb}{0.21,0.21,0.21}
\definecolor{gray22}{rgb}{0.22,0.22,0.22}
\definecolor{gray23}{rgb}{0.23,0.23,0.23}
\definecolor{gray24}{rgb}{0.24,0.24,0.24}
\definecolor{gray25}{rgb}{0.25,0.25,0.25}
\definecolor{gray26}{rgb}{0.26,0.26,0.26}
\definecolor{gray27}{rgb}{0.27,0.27,0.27}
\definecolor{gray28}{rgb}{0.28,0.28,0.28}
\definecolor{gray29}{rgb}{0.29,0.29,0.29}
\definecolor{gray2}{rgb}{0.02,0.02,0.02}
\definecolor{gray30}{rgb}{0.30,0.30,0.30}
\definecolor{gray31}{rgb}{0.31,0.31,0.31}
\definecolor{gray32}{rgb}{0.32,0.32,0.32}
\definecolor{gray33}{rgb}{0.33,0.33,0.33}
\definecolor{gray34}{rgb}{0.34,0.34,0.34}
\definecolor{gray35}{rgb}{0.35,0.35,0.35}
\definecolor{gray36}{rgb}{0.36,0.36,0.36}
\definecolor{gray37}{rgb}{0.37,0.37,0.37}
\definecolor{gray38}{rgb}{0.38,0.38,0.38}
\definecolor{gray39}{rgb}{0.39,0.39,0.39}
\definecolor{gray3}{rgb}{0.03,0.03,0.03}
\definecolor{gray40}{rgb}{0.40,0.40,0.40}
\definecolor{gray41}{rgb}{0.41,0.41,0.41}
\definecolor{gray42}{rgb}{0.42,0.42,0.42}
\definecolor{gray43}{rgb}{0.43,0.43,0.43}
\definecolor{gray44}{rgb}{0.44,0.44,0.44}
\definecolor{gray45}{rgb}{0.45,0.45,0.45}
\definecolor{gray46}{rgb}{0.46,0.46,0.46}
\definecolor{gray47}{rgb}{0.47,0.47,0.47}
\definecolor{gray48}{rgb}{0.48,0.48,0.48}
\definecolor{gray49}{rgb}{0.49,0.49,0.49}
\definecolor{gray4}{rgb}{0.04,0.04,0.04}
\definecolor{gray50}{rgb}{0.50,0.50,0.50}
\definecolor{gray51}{rgb}{0.51,0.51,0.51}
\definecolor{gray52}{rgb}{0.52,0.52,0.52}
\definecolor{gray53}{rgb}{0.53,0.53,0.53}
\definecolor{gray54}{rgb}{0.54,0.54,0.54}
\definecolor{gray55}{rgb}{0.55,0.55,0.55}
\definecolor{gray56}{rgb}{0.56,0.56,0.56}
\definecolor{gray57}{rgb}{0.57,0.57,0.57}
\definecolor{gray58}{rgb}{0.58,0.58,0.58}
\definecolor{gray59}{rgb}{0.59,0.59,0.59}
\definecolor{gray5}{rgb}{0.05,0.05,0.05}
\definecolor{gray60}{rgb}{0.60,0.60,0.60}
\definecolor{gray61}{rgb}{0.61,0.61,0.61}
\definecolor{gray62}{rgb}{0.62,0.62,0.62}
\definecolor{gray63}{rgb}{0.63,0.63,0.63}
\definecolor{gray64}{rgb}{0.64,0.64,0.64}
\definecolor{gray65}{rgb}{0.65,0.65,0.65}
\definecolor{gray66}{rgb}{0.66,0.66,0.66}
\definecolor{gray67}{rgb}{0.67,0.67,0.67}
\definecolor{gray68}{rgb}{0.68,0.68,0.68}
\definecolor{gray69}{rgb}{0.69,0.69,0.69}
\definecolor{gray6}{rgb}{0.06,0.06,0.06}
\definecolor{gray70}{rgb}{0.70,0.70,0.70}
\definecolor{gray71}{rgb}{0.71,0.71,0.71}
\definecolor{gray72}{rgb}{0.72,0.72,0.72}
\definecolor{gray73}{rgb}{0.73,0.73,0.73}
\definecolor{gray74}{rgb}{0.74,0.74,0.74}
\definecolor{gray75}{rgb}{0.75,0.75,0.75}
\definecolor{gray76}{rgb}{0.76,0.76,0.76}
\definecolor{gray77}{rgb}{0.77,0.77,0.77}
\definecolor{gray78}{rgb}{0.78,0.78,0.78}
\definecolor{gray79}{rgb}{0.79,0.79,0.79}
\definecolor{gray7}{rgb}{0.07,0.07,0.07}
\definecolor{gray80}{rgb}{0.80,0.80,0.80}
\definecolor{gray81}{rgb}{0.81,0.81,0.81}
\definecolor{gray82}{rgb}{0.82,0.82,0.82}
\definecolor{gray83}{rgb}{0.83,0.83,0.83}
\definecolor{gray84}{rgb}{0.84,0.84,0.84}
\definecolor{gray85}{rgb}{0.85,0.85,0.85}
\definecolor{gray86}{rgb}{0.86,0.86,0.86}
\definecolor{gray87}{rgb}{0.87,0.87,0.87}
\definecolor{gray88}{rgb}{0.88,0.88,0.88}
\definecolor{gray89}{rgb}{0.89,0.89,0.89}
\definecolor{gray8}{rgb}{0.08,0.08,0.08}
\definecolor{gray90}{rgb}{0.90,0.90,0.90}
\definecolor{gray91}{rgb}{0.91,0.91,0.91}
\definecolor{gray92}{rgb}{0.92,0.92,0.92}
\definecolor{gray93}{rgb}{0.93,0.93,0.93}
\definecolor{gray94}{rgb}{0.94,0.94,0.94}
\definecolor{gray95}{rgb}{0.95,0.95,0.95}
\definecolor{gray96}{rgb}{0.96,0.96,0.96}
\definecolor{gray97}{rgb}{0.97,0.97,0.97}
\definecolor{gray98}{rgb}{0.98,0.98,0.98}
\definecolor{gray99}{rgb}{0.99,0.99,0.99}
\definecolor{gray9}{rgb}{0.09,0.09,0.09}
\definecolor{gray}{rgb}{0.75,0.75,0.75}
\definecolor{green1}{rgb}{0.00,1.00,0.00}
\definecolor{green2}{rgb}{0.00,0.93,0.00}
\definecolor{green3}{rgb}{0.00,0.80,0.00}
\definecolor{green4}{rgb}{0.00,0.55,0.00}
\definecolor{greenyellow}{rgb}{0.68,1.00,0.18}
\definecolor{green}{rgb}{0.00,1.00,0.00}
\definecolor{grey0}{rgb}{0.00,0.00,0.00}
\definecolor{grey100}{rgb}{1.00,1.00,1.00}
\definecolor{grey10}{rgb}{0.10,0.10,0.10}
\definecolor{grey11}{rgb}{0.11,0.11,0.11}
\definecolor{grey12}{rgb}{0.12,0.12,0.12}
\definecolor{grey13}{rgb}{0.13,0.13,0.13}
\definecolor{grey14}{rgb}{0.14,0.14,0.14}
\definecolor{grey15}{rgb}{0.15,0.15,0.15}
\definecolor{grey16}{rgb}{0.16,0.16,0.16}
\definecolor{grey17}{rgb}{0.17,0.17,0.17}
\definecolor{grey18}{rgb}{0.18,0.18,0.18}
\definecolor{grey19}{rgb}{0.19,0.19,0.19}
\definecolor{grey1}{rgb}{0.01,0.01,0.01}
\definecolor{grey20}{rgb}{0.20,0.20,0.20}
\definecolor{grey21}{rgb}{0.21,0.21,0.21}
\definecolor{grey22}{rgb}{0.22,0.22,0.22}
\definecolor{grey23}{rgb}{0.23,0.23,0.23}
\definecolor{grey24}{rgb}{0.24,0.24,0.24}
\definecolor{grey25}{rgb}{0.25,0.25,0.25}
\definecolor{grey26}{rgb}{0.26,0.26,0.26}
\definecolor{grey27}{rgb}{0.27,0.27,0.27}
\definecolor{grey28}{rgb}{0.28,0.28,0.28}
\definecolor{grey29}{rgb}{0.29,0.29,0.29}
\definecolor{grey2}{rgb}{0.02,0.02,0.02}
\definecolor{grey30}{rgb}{0.30,0.30,0.30}
\definecolor{grey31}{rgb}{0.31,0.31,0.31}
\definecolor{grey32}{rgb}{0.32,0.32,0.32}
\definecolor{grey33}{rgb}{0.33,0.33,0.33}
\definecolor{grey34}{rgb}{0.34,0.34,0.34}
\definecolor{grey35}{rgb}{0.35,0.35,0.35}
\definecolor{grey36}{rgb}{0.36,0.36,0.36}
\definecolor{grey37}{rgb}{0.37,0.37,0.37}
\definecolor{grey38}{rgb}{0.38,0.38,0.38}
\definecolor{grey39}{rgb}{0.39,0.39,0.39}
\definecolor{grey3}{rgb}{0.03,0.03,0.03}
\definecolor{grey40}{rgb}{0.40,0.40,0.40}
\definecolor{grey41}{rgb}{0.41,0.41,0.41}
\definecolor{grey42}{rgb}{0.42,0.42,0.42}
\definecolor{grey43}{rgb}{0.43,0.43,0.43}
\definecolor{grey44}{rgb}{0.44,0.44,0.44}
\definecolor{grey45}{rgb}{0.45,0.45,0.45}
\definecolor{grey46}{rgb}{0.46,0.46,0.46}
\definecolor{grey47}{rgb}{0.47,0.47,0.47}
\definecolor{grey48}{rgb}{0.48,0.48,0.48}
\definecolor{grey49}{rgb}{0.49,0.49,0.49}
\definecolor{grey4}{rgb}{0.04,0.04,0.04}
\definecolor{grey50}{rgb}{0.50,0.50,0.50}
\definecolor{grey51}{rgb}{0.51,0.51,0.51}
\definecolor{grey52}{rgb}{0.52,0.52,0.52}
\definecolor{grey53}{rgb}{0.53,0.53,0.53}
\definecolor{grey54}{rgb}{0.54,0.54,0.54}
\definecolor{grey55}{rgb}{0.55,0.55,0.55}
\definecolor{grey56}{rgb}{0.56,0.56,0.56}
\definecolor{grey57}{rgb}{0.57,0.57,0.57}
\definecolor{grey58}{rgb}{0.58,0.58,0.58}
\definecolor{grey59}{rgb}{0.59,0.59,0.59}
\definecolor{grey5}{rgb}{0.05,0.05,0.05}
\definecolor{grey60}{rgb}{0.60,0.60,0.60}
\definecolor{grey61}{rgb}{0.61,0.61,0.61}
\definecolor{grey62}{rgb}{0.62,0.62,0.62}
\definecolor{grey63}{rgb}{0.63,0.63,0.63}
\definecolor{grey64}{rgb}{0.64,0.64,0.64}
\definecolor{grey65}{rgb}{0.65,0.65,0.65}
\definecolor{grey66}{rgb}{0.66,0.66,0.66}
\definecolor{grey67}{rgb}{0.67,0.67,0.67}
\definecolor{grey68}{rgb}{0.68,0.68,0.68}
\definecolor{grey69}{rgb}{0.69,0.69,0.69}
\definecolor{grey6}{rgb}{0.06,0.06,0.06}
\definecolor{grey70}{rgb}{0.70,0.70,0.70}
\definecolor{grey71}{rgb}{0.71,0.71,0.71}
\definecolor{grey72}{rgb}{0.72,0.72,0.72}
\definecolor{grey73}{rgb}{0.73,0.73,0.73}
\definecolor{grey74}{rgb}{0.74,0.74,0.74}
\definecolor{grey75}{rgb}{0.75,0.75,0.75}
\definecolor{grey76}{rgb}{0.76,0.76,0.76}
\definecolor{grey77}{rgb}{0.77,0.77,0.77}
\definecolor{grey78}{rgb}{0.78,0.78,0.78}
\definecolor{grey79}{rgb}{0.79,0.79,0.79}
\definecolor{grey7}{rgb}{0.07,0.07,0.07}
\definecolor{grey80}{rgb}{0.80,0.80,0.80}
\definecolor{grey81}{rgb}{0.81,0.81,0.81}
\definecolor{grey82}{rgb}{0.82,0.82,0.82}
\definecolor{grey83}{rgb}{0.83,0.83,0.83}
\definecolor{grey84}{rgb}{0.84,0.84,0.84}
\definecolor{grey85}{rgb}{0.85,0.85,0.85}
\definecolor{grey86}{rgb}{0.86,0.86,0.86}
\definecolor{grey87}{rgb}{0.87,0.87,0.87}
\definecolor{grey88}{rgb}{0.88,0.88,0.88}
\definecolor{grey89}{rgb}{0.89,0.89,0.89}
\definecolor{grey8}{rgb}{0.08,0.08,0.08}
\definecolor{grey90}{rgb}{0.90,0.90,0.90}
\definecolor{grey91}{rgb}{0.91,0.91,0.91}
\definecolor{grey92}{rgb}{0.92,0.92,0.92}
\definecolor{grey93}{rgb}{0.93,0.93,0.93}
\definecolor{grey94}{rgb}{0.94,0.94,0.94}
\definecolor{grey95}{rgb}{0.95,0.95,0.95}
\definecolor{grey96}{rgb}{0.96,0.96,0.96}
\definecolor{grey97}{rgb}{0.97,0.97,0.97}
\definecolor{grey98}{rgb}{0.98,0.98,0.98}
\definecolor{grey99}{rgb}{0.99,0.99,0.99}
\definecolor{grey9}{rgb}{0.09,0.09,0.09}
\definecolor{grey}{rgb}{0.75,0.75,0.75}
\definecolor{honeydew1}{rgb}{0.94,1.00,0.94}
\definecolor{honeydew2}{rgb}{0.88,0.93,0.88}
\definecolor{honeydew3}{rgb}{0.76,0.80,0.76}
\definecolor{honeydew4}{rgb}{0.51,0.55,0.51}
\definecolor{honeydew}{rgb}{0.94,1.00,0.94}
\definecolor{hotpink}{rgb}{1.00,0.41,0.71}
\definecolor{indianred}{rgb}{0.80,0.36,0.36}
\definecolor{ivory1}{rgb}{1.00,1.00,0.94}
\definecolor{ivory2}{rgb}{0.93,0.93,0.88}
\definecolor{ivory3}{rgb}{0.80,0.80,0.76}
\definecolor{ivory4}{rgb}{0.55,0.55,0.51}
\definecolor{ivory}{rgb}{1.00,1.00,0.94}
\definecolor{khaki1}{rgb}{1.00,0.96,0.56}
\definecolor{khaki2}{rgb}{0.93,0.90,0.52}
\definecolor{khaki3}{rgb}{0.80,0.78,0.45}
\definecolor{khaki4}{rgb}{0.55,0.53,0.31}
\definecolor{khaki}{rgb}{0.94,0.90,0.55}
\definecolor{lavenderblush}{rgb}{1.00,0.94,0.96}
\definecolor{lavender}{rgb}{0.90,0.90,0.98}
\definecolor{lawngreen}{rgb}{0.49,0.99,0.00}
\definecolor{lemonchiffon}{rgb}{1.00,0.98,0.80}
\definecolor{lightblue}{rgb}{0.68,0.85,0.90}
\definecolor{lightcoral}{rgb}{0.94,0.50,0.50}
\definecolor{lightcyan}{rgb}{0.88,1.00,1.00}
\definecolor{lightgoldenrod}{rgb}{0.93,0.87,0.51}
\definecolor{lightgoldenrod}{rgb}{0.98,0.98,0.82}
\definecolor{lightgray}{rgb}{0.83,0.83,0.83}
\definecolor{lightgreen}{rgb}{0.56,0.93,0.56}
\definecolor{lightgrey}{rgb}{0.83,0.83,0.83}
\definecolor{lightpink}{rgb}{1.00,0.71,0.76}
\definecolor{lightsalmon}{rgb}{1.00,0.63,0.48}
\definecolor{lightsea}{rgb}{0.13,0.70,0.67}
\definecolor{lightsky}{rgb}{0.53,0.81,0.98}
\definecolor{lightslate}{rgb}{0.47,0.53,0.60}
\definecolor{lightslate}{rgb}{0.47,0.53,0.60}
\definecolor{lightslate}{rgb}{0.52,0.44,1.00}
\definecolor{lightsteel}{rgb}{0.69,0.77,0.87}
\definecolor{lightyellow}{rgb}{1.00,1.00,0.88}
\definecolor{limegreen}{rgb}{0.20,0.80,0.20}
\definecolor{linen}{rgb}{0.98,0.94,0.90}
\definecolor{magenta1}{rgb}{1.00,0.00,1.00}
\definecolor{magenta2}{rgb}{0.93,0.00,0.93}
\definecolor{magenta3}{rgb}{0.80,0.00,0.80}
\definecolor{magenta4}{rgb}{0.55,0.00,0.55}
\definecolor{magenta}{rgb}{1.00,0.00,1.00}
\definecolor{maroon1}{rgb}{1.00,0.20,0.70}
\definecolor{maroon2}{rgb}{0.93,0.19,0.65}
\definecolor{maroon3}{rgb}{0.80,0.16,0.56}
\definecolor{maroon4}{rgb}{0.55,0.11,0.38}
\definecolor{maroon}{rgb}{0.69,0.19,0.38}
\definecolor{mediumaquamarine}{rgb}{0.40,0.80,0.67}
\definecolor{mediumblue}{rgb}{0.00,0.00,0.80}
\definecolor{mediumorchid}{rgb}{0.73,0.33,0.83}
\definecolor{mediumpurple}{rgb}{0.58,0.44,0.86}
\definecolor{mediumsea}{rgb}{0.24,0.70,0.44}
\definecolor{mediumslate}{rgb}{0.48,0.41,0.93}
\definecolor{mediumspring}{rgb}{0.00,0.98,0.60}
\definecolor{mediumturquoise}{rgb}{0.28,0.82,0.80}
\definecolor{mediumviolet}{rgb}{0.78,0.08,0.52}
\definecolor{midnightblue}{rgb}{0.10,0.10,0.44}
\definecolor{mintcream}{rgb}{0.96,1.00,0.98}
\definecolor{mistyrose}{rgb}{1.00,0.89,0.88}
\definecolor{moccasin}{rgb}{1.00,0.89,0.71}
\definecolor{navajowhite}{rgb}{1.00,0.87,0.68}
\definecolor{navyblue}{rgb}{0.00,0.00,0.50}
\definecolor{navy}{rgb}{0.00,0.00,0.50}
\definecolor{oldlace}{rgb}{0.99,0.96,0.90}
\definecolor{olivedrab}{rgb}{0.42,0.56,0.14}
\definecolor{orange1}{rgb}{1.00,0.65,0.00}
\definecolor{orange2}{rgb}{0.93,0.60,0.00}
\definecolor{orange3}{rgb}{0.80,0.52,0.00}
\definecolor{orange4}{rgb}{0.55,0.35,0.00}
\definecolor{orangered}{rgb}{1.00,0.27,0.00}
\definecolor{orange}{rgb}{1.00,0.65,0.00}
\definecolor{orchid1}{rgb}{1.00,0.51,0.98}
\definecolor{orchid2}{rgb}{0.93,0.48,0.91}
\definecolor{orchid3}{rgb}{0.80,0.41,0.79}
\definecolor{orchid4}{rgb}{0.55,0.28,0.54}
\definecolor{orchid}{rgb}{0.85,0.44,0.84}
\definecolor{palegoldenrod}{rgb}{0.93,0.91,0.67}
\definecolor{palegreen}{rgb}{0.60,0.98,0.60}
\definecolor{paleturquoise}{rgb}{0.69,0.93,0.93}
\definecolor{paleviolet}{rgb}{0.86,0.44,0.58}
\definecolor{papayawhip}{rgb}{1.00,0.94,0.84}
\definecolor{peachpuff}{rgb}{1.00,0.85,0.73}
\definecolor{peru}{rgb}{0.80,0.52,0.25}
\definecolor{pink1}{rgb}{1.00,0.71,0.77}
\definecolor{pink2}{rgb}{0.93,0.66,0.72}
\definecolor{pink3}{rgb}{0.80,0.57,0.62}
\definecolor{pink4}{rgb}{0.55,0.39,0.42}
\definecolor{pink}{rgb}{1.00,0.75,0.80}
\definecolor{plum1}{rgb}{1.00,0.73,1.00}
\definecolor{plum2}{rgb}{0.93,0.68,0.93}
\definecolor{plum3}{rgb}{0.80,0.59,0.80}
\definecolor{plum4}{rgb}{0.55,0.40,0.55}
\definecolor{plum}{rgb}{0.87,0.63,0.87}
\definecolor{powderblue}{rgb}{0.69,0.88,0.90}
\definecolor{purple1}{rgb}{0.61,0.19,1.00}
\definecolor{purple2}{rgb}{0.57,0.17,0.93}
\definecolor{purple3}{rgb}{0.49,0.15,0.80}
\definecolor{purple4}{rgb}{0.33,0.10,0.55}
\definecolor{purple}{rgb}{0.63,0.13,0.94}
\definecolor{red1}{rgb}{1.00,0.00,0.00}
\definecolor{red2}{rgb}{0.93,0.00,0.00}
\definecolor{red3}{rgb}{0.80,0.00,0.00}
\definecolor{red4}{rgb}{0.55,0.00,0.00}
\definecolor{red}{rgb}{1.00,0.00,0.00}
\definecolor{rosybrown}{rgb}{0.74,0.56,0.56}
\definecolor{royalblue}{rgb}{0.25,0.41,0.88}
\definecolor{saddlebrown}{rgb}{0.55,0.27,0.07}
\definecolor{salmon1}{rgb}{1.00,0.55,0.41}
\definecolor{salmon2}{rgb}{0.93,0.51,0.38}
\definecolor{salmon3}{rgb}{0.80,0.44,0.33}
\definecolor{salmon4}{rgb}{0.55,0.30,0.22}
\definecolor{salmon}{rgb}{0.98,0.50,0.45}
\definecolor{sandybrown}{rgb}{0.96,0.64,0.38}
\definecolor{seagreen}{rgb}{0.18,0.55,0.34}
\definecolor{seashell1}{rgb}{1.00,0.96,0.93}
\definecolor{seashell2}{rgb}{0.93,0.90,0.87}
\definecolor{seashell3}{rgb}{0.80,0.77,0.75}
\definecolor{seashell4}{rgb}{0.55,0.53,0.51}
\definecolor{seashell}{rgb}{1.00,0.96,0.93}
\definecolor{sienna1}{rgb}{1.00,0.51,0.28}
\definecolor{sienna2}{rgb}{0.93,0.47,0.26}
\definecolor{sienna3}{rgb}{0.80,0.41,0.22}
\definecolor{sienna4}{rgb}{0.55,0.28,0.15}
\definecolor{sienna}{rgb}{0.63,0.32,0.18}
\definecolor{skyblue}{rgb}{0.53,0.81,0.92}
\definecolor{slateblue}{rgb}{0.42,0.35,0.80}
\definecolor{slategray}{rgb}{0.44,0.50,0.56}
\definecolor{slategrey}{rgb}{0.44,0.50,0.56}
\definecolor{snow1}{rgb}{1.00,0.98,0.98}
\definecolor{snow2}{rgb}{0.93,0.91,0.91}
\definecolor{snow3}{rgb}{0.80,0.79,0.79}
\definecolor{snow4}{rgb}{0.55,0.54,0.54}
\definecolor{snow}{rgb}{1.00,0.98,0.98}
\definecolor{springgreen}{rgb}{0.00,1.00,0.50}
\definecolor{steelblue}{rgb}{0.27,0.51,0.71}
\definecolor{tan1}{rgb}{1.00,0.65,0.31}
\definecolor{tan2}{rgb}{0.93,0.60,0.29}
\definecolor{tan3}{rgb}{0.80,0.52,0.25}
\definecolor{tan4}{rgb}{0.55,0.35,0.17}
\definecolor{tan}{rgb}{0.82,0.71,0.55}
\definecolor{thistle1}{rgb}{1.00,0.88,1.00}
\definecolor{thistle2}{rgb}{0.93,0.82,0.93}
\definecolor{thistle3}{rgb}{0.80,0.71,0.80}
\definecolor{thistle4}{rgb}{0.55,0.48,0.55}
\definecolor{thistle}{rgb}{0.85,0.75,0.85}
\definecolor{tomato1}{rgb}{1.00,0.39,0.28}
\definecolor{tomato2}{rgb}{0.93,0.36,0.26}
\definecolor{tomato3}{rgb}{0.80,0.31,0.22}
\definecolor{tomato4}{rgb}{0.55,0.21,0.15}
\definecolor{tomato}{rgb}{1.00,0.39,0.28}
\definecolor{turquoise1}{rgb}{0.00,0.96,1.00}
\definecolor{turquoise2}{rgb}{0.00,0.90,0.93}
\definecolor{turquoise3}{rgb}{0.00,0.77,0.80}
\definecolor{turquoise4}{rgb}{0.00,0.53,0.55}
\definecolor{turquoise}{rgb}{0.25,0.88,0.82}
\definecolor{violetred}{rgb}{0.82,0.13,0.56}
\definecolor{violet}{rgb}{0.93,0.51,0.93}
\definecolor{wheat1}{rgb}{1.00,0.91,0.73}
\definecolor{wheat2}{rgb}{0.93,0.85,0.68}
\definecolor{wheat3}{rgb}{0.80,0.73,0.59}
\definecolor{wheat4}{rgb}{0.55,0.49,0.40}
\definecolor{wheat}{rgb}{0.96,0.87,0.70}
\definecolor{whitesmoke}{rgb}{0.96,0.96,0.96}
\definecolor{white}{rgb}{1.00,1.00,1.00}
\definecolor{yellow1}{rgb}{1.00,1.00,0.00}
\definecolor{yellow2}{rgb}{0.93,0.93,0.00}
\definecolor{yellow3}{rgb}{0.80,0.80,0.00}
\definecolor{yellow4}{rgb}{0.55,0.55,0.00}
\definecolor{yellowgreen}{rgb}{0.60,0.80,0.20}
\definecolor{yellow}{rgb}{1.00,1.00,0.00}

\usepackage{ulem}       
\begin{document}        


\maketitle
\label{firstpage}

\begin{abstract} Following the study of Darg et al. (2009; hereafter D09a) we explore the environments, optical colours, stellar masses, star formation and AGN activity in a sample of 3003 pairs of merging galaxies drawn from the SDSS using {\color{black}visual classifications} from the Galaxy Zoo project. While D09a found that the spiral-to-elliptical ratio in (major) mergers appeared higher than that of the global galaxy population, no significant differences are found between the environmental distributions of mergers and a randomly selected control sample. This makes the high occurrence of spirals in mergers unlikely to be an environmental effect and must, therefore, arise from differing time-scales of detectability for spirals and ellipticals. We find that merging galaxies have a wider spread in colour than the global galaxy population, with a significant blue tail resulting from intense star formation in spiral mergers. Galaxies classed as star-forming using their emission-line properties have average star-formation rates approximately doubled by the merger process though star formation is negligibly enhanced in merging elliptical galaxies. We conclude that the internal properties of galaxies significantly affect the time-scales over which merging systems can be detected (as suggested by recent theoretical studies) which leads to spirals being `over-observed' in mergers. We also suggest that the transition mass $3\times10^{10}\mbox{M}_{\astrosun}$, noted by \citet{kauffmann1}, below which ellipticals are rare could be linked to disc survival/destruction in mergers.
\end{abstract}



\section{Introduction}

\label{intro}

{\color{black} 
The Galaxy Zoo project\footnote{www.galaxyzoo.org. The original site which produced the results for this paper is preserved at http://zoo1.galaxyzoo.org.} \citep{lintott1} has helped meet the need for extensive visual classification of galaxy morphologies in a manner that does not presuppose their spectro-photometric characteristics. Prior to this, large-scale studies of morphology from surveys like SDSS required the use of proxies in place of actual morphology (e.g. as in \citealt{bernardi}), but this then biased what their spectro-photometric characteristics {\it actually} are (\citealt{schawinski2}).

Similarly, locating mergers via structural parameters such as concentration and asymmetry (\citealt{con2003}) is problematic due to the great variety of configurations of mergers (progenitor types, impact parameters and the stage at which the system is viewed) making it difficult to define a parameter space uniquely occupied by mergers. For these reasons visual examination of images of galaxies (assuming they are at redshifts low enough for a decent resolution) is the best way to identify strongly-perturbed systems and produce merger catalogues with minimal contamination (see D09a for a more detailed discussion of merger-location techniques)}. 

In the first Galaxy Zoo mergers paper, D09a, we demonstrated how the Galaxy Zoo web-interface is able to accomplish this by creating a measure (called the {\it weighted merger-vote fraction}, $f_m\in [0,1]$; D09a Eq. 1) of how `merger-like' an SDSS image appears to be (where an image attaining $f_m=1.0$ is certain to be a merger and one with $f_m=0$ is certain not to be). The $f_m$ values were used to estimate the fraction of major mergers (where the stellar masses of the progenitors $\mbox{M}^*_1$ and $\mbox{M}^*_2$ satisfy $1/3<\mbox{M}_1^*/\mbox{M}_2^*<3$) in the local universe to be $1-3 \times C \%$ for $M_r<-20.55$ where $C \sim 1.5$ is a correction factor for spectroscopic incompleteness. 

We found that most systems with $f_m>0.4$ can confidently be identified as mergers and used this limit to isolate 3003 merging pairs in the range $0.005<z<0.1$.\footnote{This catalogue is GZM1: Galaxy Zoo Mergers 1.} All 3003 pairs were then visually examined in order to assign morphologies to the individual galaxies in each merger. From this, we found that the spiral-to-elliptical ratio ($N_s/N_e$) in merging systems was higher in our sample ($N_s/N_e \ga3$ for $f_m>0.4$) compared to the global galaxy population ($N_s/N_e\sim1.5$) which was determined using the statistical corrections of \citet{bamford} for all Galaxy Zoo morphologies. We argued that the observed spiral excess is real for major mergers, i.e., unlikely to be compensated by an excess of ellipticals in the range $f_m<0.4$.

It is not a surprise that $N_s/N_e$ should differ between mergers and the global population. The probability of {\it observing} a merger at any general time is proportional to the likelihood of it merging with another galaxy (which depends on its environment) and the time-scale over which the merger is detectable (which depends on the internal properties of the progenitors). {\color{black} Since spirals differ from ellipticals in both environment \citep{dressler} and their internal properties, it is not unreasonable to expect $N_s/N_e$ in merger observations to deviate from that of the complete galaxy population.} 

Recent simulations by \citet{lotz3} examined the time-scales of detectability for the merger-detection techniques of `close-pairs' and combinations of non-parametric quantities (namely C, A, G \& M$_{20}$; see \citealp{con2008} and D09a for discussion). These simulations found that the internal properties of the progenitors significantly affect the time-scales over which real systems would have been flagged as merging. The study also found that the physical factors that had the greatest affect on the time-scales of detectability were the gas-fractions of the progenitors, their pericentric separation and their relative orientation. Conversely, the choice of system mass and supernova-feedback prescription only slightly affected the time-scales of detectability.  

This would suggest that relative gas content is the most {\it morphology-specific} factor affecting the time-scales of detectability since spirals are relatively gas-rich compared to ellipticals (orientation and pericentric separation, on the other hand, are likely to be independent of morphology).

The internal properties of merging galaxies are also important in so far as they determine the morphological outcome of interacting galaxies. The abundance of minor mergers in the universe (e.g. \citealt{woods2}) must mean that disc galaxies survive substantial numbers of minor mergers. Furthermore, given the estimate by \citet{con2008} that, on average, a galaxy in the mass range $\mbox{M}^*>10^{10}\mbox{M}_{\astrosun}$ will have undergone $4.3^{+0.8}_{-0.8}$ major mergers since $z \sim 3$, the number of disc galaxies in the local universe should appear sizably reduced from that observed {\it unless} they too can survive major mergers (\citealt{hopkins}; \citealt{hopkins2}). Disc survival in major mergers is therefore likely to become an important principle in galaxy evolution that we aim to shed light upon in this work. The study of \citealp{hopkins} places great emphasis on the gas-to-stellar-mass ratio in progenitor discs which implicates the importance of feedback processes in mergers (supernovae and AGN) in so far as they can help retain gas at large radii whose angular momentum generates disc reformation after dynamical relaxation. 

Star-formation histories are therefore important to study in mergers with respect to gas-retention in discs. More broadly, several matters remain unsolved regarding galactic star-formation histories (\citealt{ellis}; \citealt{kaviraj2}) in which mergers play an important role as they are thought to directly trigger star formation (\citealt{schweizer2}; \citealt{li}), generate (Ultra-) Luminous-Infra-Red-Galaxy activity (\citealt{sanders}; \citealt{kaviraj}; \citealt{genzel}) and bring about the formation of clusters (\citealt{zepf}; \citealt{schweizer3}). Empirical confirmation of the extent to which mergers are able to affect the luminosity function is thus an important task. 

Mergers can also affect star formation and disc dynamics in so far as they trigger AGN activity. This is thought to be a natural consequence of the angular-momentum loss that can take place in galactic interactions allowing the infall of gas (\citealt{kewley1}) that fuels the central super-massive black hole (\citealt{somerville}; \citealt{jog2}). AGN feedback then controls further infall and cooling of gas leading to reduced star formation (\citealt{schawinski1}; \citealt{schawinski2}; \citealt{khalatyan}; \citealt{schawinski5}). Studies showing star-burst activity within AGN galaxies have suggested mergers as the causal mechanism (e.g. \citealt{storchi}, \citealt{kauffmann2}). We test some of these claims by inspecting the star-formation and AGN signatures in our mergers using emission-line diagnostics.

In \S\ref{samples} we recap the catalogue conventions described in detail in D09a and describe the construction of our control sample. To disentangle the role of environment from internal properties on a galaxy's probability of being observed in a merger, we begin our investigation with a study of the environmental distributions of our samples in \S\ref{enviro} which we model by a single degree of freedom. The internal properties of galaxies require many more degrees of freedom to fully capture their dynamics. We approach this task by first examining the photometry of our samples (\S\ref{colour}). In section \S\ref{masses} we examine the stellar-mass distributions of our samples and how these correlate with morphologies and colours in merging systems. In \S\ref{agn_stuff} we perform spectral-line diagnostics to determine the main ionisation sources (or lack of) in our samples accompanied by star-formation-rate estimates. We summarise and discuss our results in \S\ref{conc}.

\section{The Merger and Control Samples}
\label{samples}

\subsection{Sample Morphologies}
\label{morphs}
\begin{figure}
\includegraphics[width=84mm]{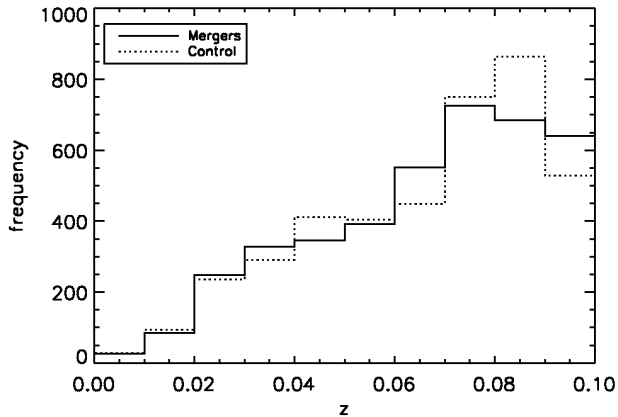}
\caption[Figure 6b]{The redshift distributions of the merger and control samples appear roughly the same within counting errors as expected. Any distance-dependent aperture or deblending biases should affect both samples equally.}	
\label{hist_redshift}
\end{figure}

The construction of the merging-pairs catalogue GZM1 is described in detail in D09a. For all 3003 merging pairs, we assigned one of the following morphologies to the constituent galaxies: `Elliptical' (E), `Spiral' (S), `Unclear but most probably Elliptical' (EU) and `Unclear but most probably a Spiral' (SU). 

We then created a control sample of randomly selected spectral galaxies from the SDSS DR6 catalogue. It is from the same redshift range ($0.005<z<0.1$) so that the control and merger galaxies should have similar redshift distributions (qualitatively confirmed by Figure \ref{hist_redshift}). The random nature of the selection of control objects makes them a reasonable representation of the global galaxy population whose properties can be compared and contrasted to those of our merger sample.\footnote{{\color{black} The control sample is arbitrarily large; we select the same number of control galaxies as are being used for the mergers depending on the specifics of the investigation. For example, if we volume-limit our merger catalogue to get a sample with $x$ galaxies, we compare it to a volume-limited sample of $x$ control galaxies, etc.}} 

All calculations carried out in this paper on the merger objects are carried out in the same way for the control sample. The {\it only} difference is that on the few occasions where we split the control sample into spiral and elliptical categories we use the following criteria: a control galaxy is a spiral if the GZ weighted-spiral-vote fraction $f_s$ is greater than its weighted-elliptical-vote fraction $f_e$ (these are direct analogues to $f_m$ but for spirals and ellipticals respectively; D09a, Eq. 1) and with a minimum absolute difference of 0.1. Similarly, a control galaxy is taken to be an elliptical if $f_e>f_s+0.1$. Control galaxies with $\left|f_e-f_s \right|<0.1$ are not used when comparing morphologies to the properties of merging and non-merging galaxies since they are too hard to distinguish. It is important to note that since the merger and control morphologies were determined differently (mergers by DWD) they should only be taken as a rough guide.\footnote{{\color{black}In particular we find in \S \ref{colour} that the control ellipticals are bluer than the merger ellipticals and this will mean that stellar mass estimates will be typically lower for the control sample. It is difficult to disentangle whether this extra `blueness' in ellipticals is physical or just a selection effect. This problem will be overcome by the Galaxy Zoo Two project.}} 

\subsection{Assigning Merger Stages}
\label{vis_stage}
\begin{figure}
\begin{center}
	\includegraphics[width=84mm]{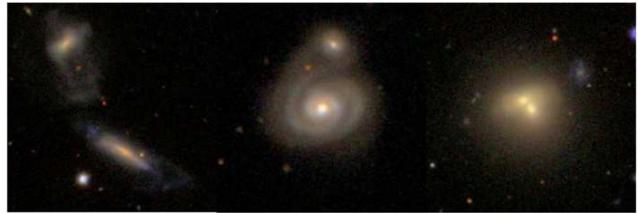}
	\caption[Figure 6b]{Example images of the visually-assigned `stage' categories: (left) `separated', (centre) `interacting' \& (right) `approaching post-merger.' Of our 3003 merging pairs, 167 ($\sim6\%$), 2526 ($\sim84\%$) and 310 ($\sim10\%$) were assigned to these categories respectively.\footnote{Image widths: Separated $\sim100''$, Interacting $\sim70''$ \& approaching post-merger $\sim50''$.}}
	\label{stages}
\end{center}
\end{figure}


For each merger pair we also assigned a visually-chosen merger `stage.' We use three categories: `separated', `interacting' and `approaching post-merger.' The `separated' stage refers to systems classified as a merger in which there is visible space between the galaxies. The `approaching post-merger' stage refers to systems where the progenitor cores are typically within $\sim5''$ of each other on the images.\footnote{Angular scales were used on all images for visual classifications.} The `interacting' stage refers to anything in between: the galaxies have coalesced to some degree with no space visible in between but the cores have not settled to $\sim5''$ yet. Figure \ref{stages} shows examples of these stages. The stages of our 3003 merging pairs comprise of 167 ($\sim6\%$) `separated,' 2526 ($\sim84\%$) `interacting' and 310 ($\sim10\%$) `approaching post-merger' stages. The mean projected separation of the galaxy objects for these three stages are $\sim27.3$kpc, $\sim12.8$kpc and $\sim5.4$kpc respectively. The mean projected separation for all 3003 pairs is $\sim12.5$kpc.


\section{The Environment of Merging Galaxies}
\label{enviro}
\begin{figure}
	\includegraphics[width=84mm]{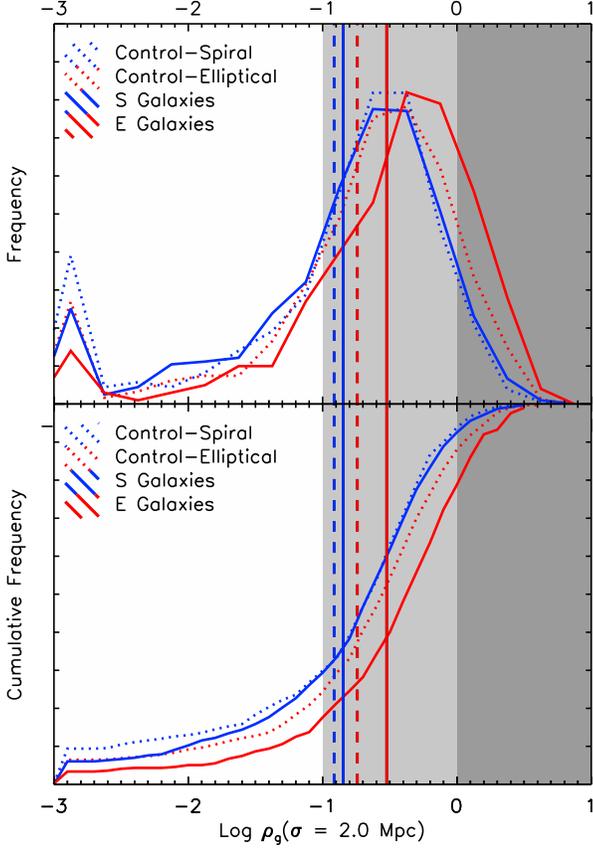}
	\caption[Figure 10]{Distributions of $\rho_g$ for merger and control populations. The distributions in the upper panel are scaled to have unitary area. We set galaxies with $\rho_g=0$ to $\rho_g=10^{-3}$ to avoid Log(0) errors (hence the spike near -3). The vertical lines mark the mean value $\left<\rho_g\right>$ for the samples. The shading in the background is an indicator of environment type. The darkest shade corresponds to $\rho_g>1.0$ - the `cluster environment.' The unshaded-white area corresponds to $\rho_g<0.1$ which we label the `field' environment.' The middle shade corresponds to `intermediate environments.'}	
	\label{plotEnviro}
\end{figure}

To parametrise environment, we employ the method of \citet{schawinski3} that takes advantage of all the spectroscopic-redshift recordings in the SDSS DR6 to obtain the dimensionless number $\rho_{g}$ (the adaptive Gaussian environment parameter) for each galaxy in our catalogue. This is a sophisticated measure of number density mapped onto $\rho_{g}\in\mathbb{R}^{+}$. The method is highly versatile and an adapted version has recently been used to locate galaxy clusters \citep{yoon}. The parameter $\rho_g(ra, dec, z, \sigma)$ starts by finding close neighbours in DR6 within an initial radius of $\sigma$ for each galaxy. We use $\sigma=2.0$Mpc following \citet{schawinski3}. It then adapts this radius depending on the initial number return in order to compensate for the ``finger-of-God'' effect and is weighted such that $\rho_g$ increases the nearer its neighbours are. Wherever we have spectra for {\it both} galaxies in the merger, we remove one of them from the calculation to avert a skewed result (as $\rho_g$ is sensitive to nearby objects).

Some example values for $\rho_g$ are useful at this point. A galaxy with the lowest value of $\rho_g=0$ has no neighbours within a $\sigma$ radius. Values up to $\rho_g=0.1$ we call the `field environment.' $\rho_g=1$ roughly corresponds to the centre of a sphere of radius 3 Mpc with ten galaxies randomly distributed within. We call this the lower limit of the `dense cluster' environment. We call galaxies with $0.1<\rho_g<1$ members of `intermediate environments.'

We plot the distribution of $\rho_g$ for our merger and control populations in Figure \ref{plotEnviro} with the background shading representing our different environments. The samples used are volume-limited (each galaxy must have $M_r<-20.55$) and we exclude `unsure' type morphologies (the distributions are near identical with them). We calculate a Kolmogorov-Smirnov (K-S) statistic for the two pairs of cumulative-frequency graphs (control-ellipticals vs. E galaxies and control-spirals vs. S galaxies). 

We find both merger and control samples are peaked in what we have called `intermediate-environments.' On average, the E galaxies occupy slightly denser environments than their control counterparts ($\left<\rho_g\right>_{\mbox{E}}-\left<\rho_g\right>_{\mbox{con.-ellip.}}\sim0.12$) with a K-S significance level of $\sim97\%$. The S galaxies appear almost unaffected and, if anything, occupy slightly denser environments than their control counterparts ($\left<\rho_g\right>_{\mbox{S}}-\left<\rho_g\right>_{\mbox{con.-spiral}}\sim0.02$ with a K-S significance level of $\sim82\%$). {\color{black} The overall distributions are virtually unaffected if we cut by mass ($>7\times 10^{10} \mbox{M}_{\astrosun}$) or if we use no mass/magnitude limit. When we combine morphologies, the mergers are, overall, in virtually identical environments as the control sample (Appendix \ref{massNenviro}). }

As mergers therefore appear to occupy similar if not slightly {\it denser} environments (environments where ellipticals are more prevalent) we can rule out the role of environment as a means to explain the high spiral-to-elliptical ratio in mergers (as reported in D09a). If anything, the tendency of mergers to occupy denser (elliptical-rich) environments ought to decrease the spiral-to-elliptical ratio in mergers. The discrepancies in the spiral-to-elliptical ratios between the merger and global populations must therefore arise from {\it longer time-scales of detectability} for mergers involving spirals than for mergers involving ellipticals. {\color{black} Thus the internal properties of galaxies that distinguish spirals from ellipticals must be such that spirals remain detectable in mergers for longer periods of time. We begin an investigation of the internal properties of galaxies by examining their colour-magnitude relations.}

\begin{figure*}
	\includegraphics[width=140mm]{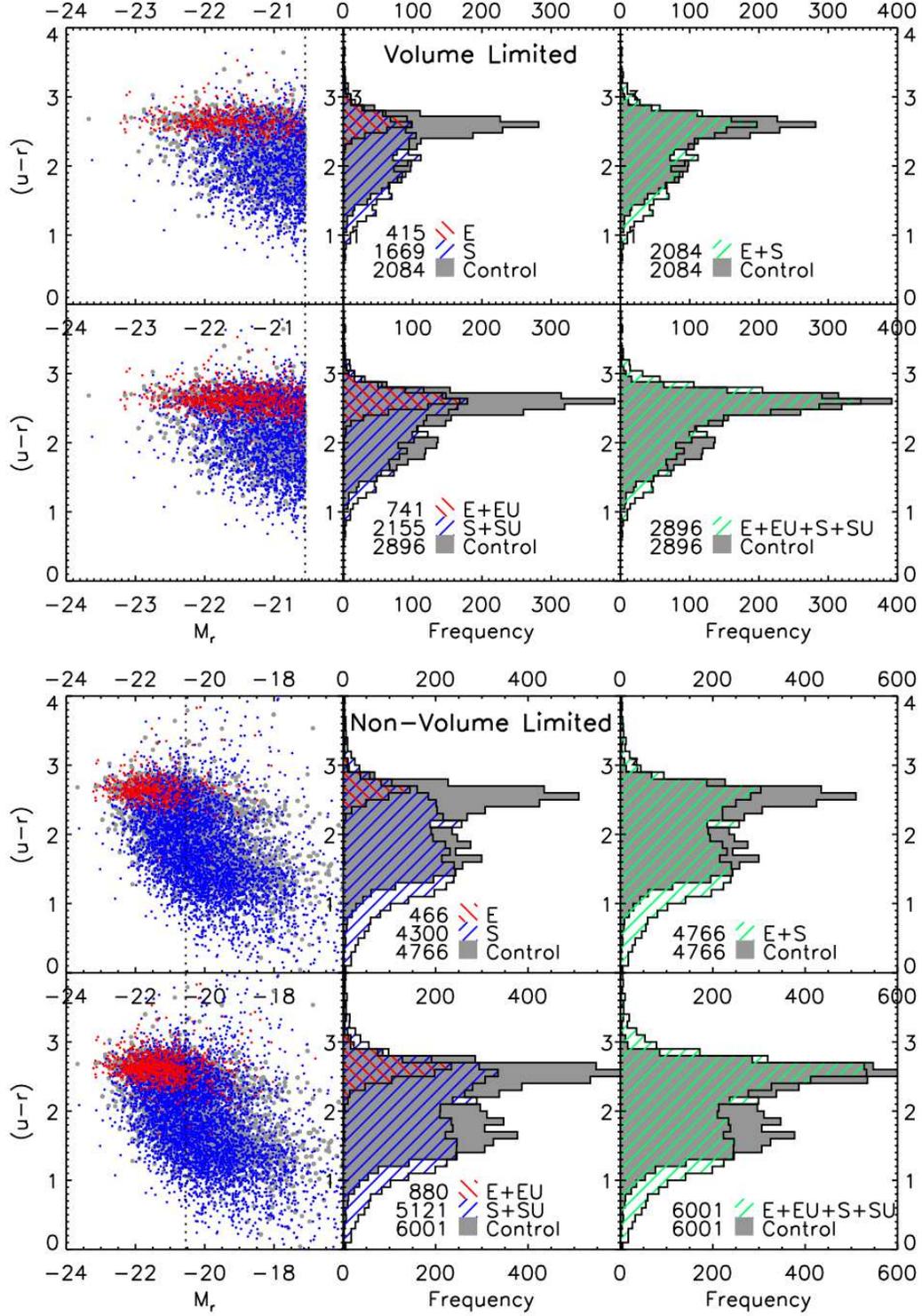}
	\caption[Figure 7]{Colour-magnitude diagrams for samples of the individual galaxies involved in our mergers (coloured) and our control sample (grey). The k-corrected rest-frame magnitude limit is $M_r<-20.55$ (broken vertical line). The upper and lower sets of figures correspond to the volume-limited and non-volume-limited samples respectively.}	
	\label{histColour}
\end{figure*}

\section{The Colours of Merging Galaxies}
\label{colour}

For all 3003 merging systems, at least one of the constituent galaxies has spectra. We use this spectral redshift to obtain k-corrected rest-frame magnitudes for both galaxies in each merger pair using the SDSS $ugriz$ model mags as inputs into the IDL routine {\texttt kcorrect 4\_1\_4} \citep{blanton}.

We examine luminosity and colour in detail for our samples (Figure \ref{histColour}) in order to gain an overview of their characteristics and to examine the qualitative effects of including the `unsure' morphologies and imposing the volume-limited constraints (D09a).

The upper and lower sets of diagrams display the volume-limited and non-volume-limited samples respectively. The volume-limited sample has both redshift and absolute magnitude bounds ($M_r<-20.55$) whereas the non-volume-limited sample only has redshift bounds. For both of these sets, the colour distributions are plotted for just the `sure' morphologies (E \& S - top rows) and then all merger morphologies (E, S, EU \& SU - bottom rows). 

All samples exhibit some bi-modality though to differing degrees. The volume-limited samples show only marginal bi-modality which is not surprising since the brighter galaxies will be dominated by galaxies in the red sequence {\color{black}(since dimmer galaxies are, on average, bluer)}. The magnitude cut of $M_r<-20.55$ removes many of the bluer, low-luminosity galaxies. We see this in the non-volume-limited diagrams which include many more (relatively-dim) galaxies that are mostly blue spirals. For both the volume-limited and non-volume-limited samples, we find that inclusion of the `unsure' morphologies makes the overall distributions more peaked in the red. Apart from this, the qualitative shapes of the distributions are roughly the same with or without the `unsure' morphologies. 

In particular we find that in all cases the mergers appear to have a higher spread in colour at both the red and blue ends compared to the control sample (see right hand columns of both the upper and lower sets of diagrams). This is in accord with early observations that `irregular' morphologies have a greater spread in colour than `regular' ones \citep{larson}. The effect is especially strong at the blue end and a natural interpretation of this is due to strong star formation induced by the merger process. We examine this possibility using emission-line diagnostics in \S \ref{sfr}. 

The slight spread at the red end might be due to increased extinction brought about by the journey of light from one galaxy core through the extra dust of the perturbing neighbour (if they lie roughly on the same line of sight). We visually examined all spirals in mergers to ensure those with $u-r>3.5$ were not red due to an edge-on view. The blue tail is more prominent for the non-volume-limited sample which, as stated, includes more low-luminosity galaxies which are almost all S or SU morphologies. This fits well with the notion that low-mass spirals have formed recently and are rich in gas and will therefore produce high specific-star-formation rates if they undergo mergers. We show in \S \ref{sfr} that low-mass spirals do in fact have the highest star-formation rates relative to their stellar mass. 

{\color{black} We also compared the colours of the control morphologies to the merger morphologies. We find that, when we use volume-limited samples, the overall means for $u-r$ are very similar between the control and merger samples. Similarly, the merger-spiral (S+SU) and control-spiral subsets have similar $u-r$ means with $\Delta \left< u-r \right> \sim 0.05$ magnitudes. However, the merger ellipticals (E+EU) have a slightly redder mean compared to the control ellipticals with $\Delta \left< u-r \right> \sim 0.15$ magnitudes. It is difficult to disentangle whether this is due to a selection effect based upon how the morphologies were selected or whether ellipticals in mergers are genuinely observed to be redder. As noted, the overall merger distributions have a more prominent red tail compared to the control distributions, and so we should not be surprised that the ellipticals in mergers are genuinely redder, only, the degree to which they are redder might be exaggerated by the morphological selection effects (see \S \ref{morphs}). This emphasises the fact that comparisons between {\it morphologies} for the merger and control samples should be taken as a rough guide only. }


\section{The Stellar Masses of Merging Galaxies}
\label{masses}
\subsection{Merger Mass Distributions}
\label{mass}
\begin{figure}
	\includegraphics[width=84mm]{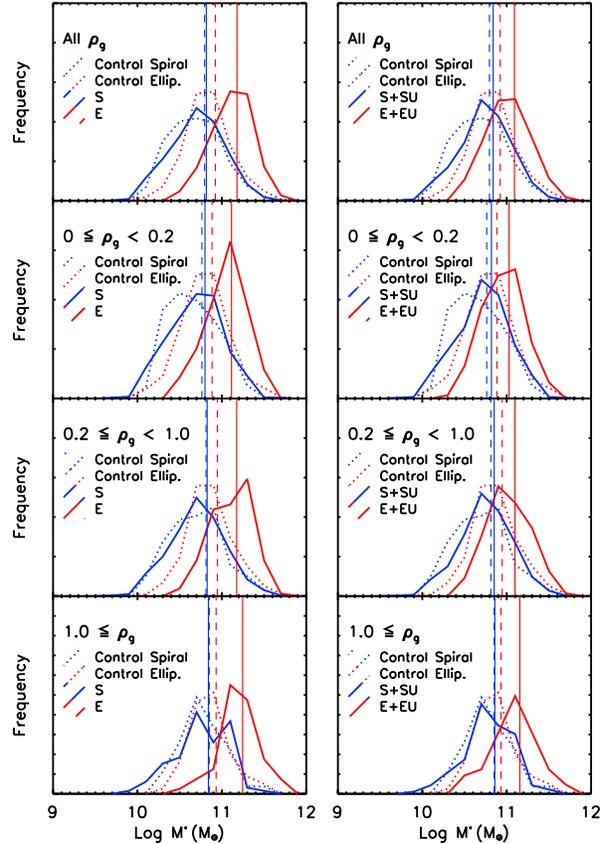}
	\caption[Figure 13]{Mass distributions of the volume-limited galaxies for differing environments corresponding to (from top to bottom) `all', `field', `intermediate' and `cluster' environments. All graphs are scaled to have unitary area. The vertical lines indicate the mean masses of the samples. {\color{black} The right-hand panels use all morphologies, the left-hand panels only the `sure' morphologies.} }	
	\label{plot_kmass}

\end{figure}

The stellar-mass estimates for our catalogues are described in detail in D09a. To recap briefly, we fit the SDSS photometry to a library of two-component star formation histories with a simple stellar population for the first burst and a variable e-folding time for the second (exponential) burst. Metallicity is fixed to solar and dust is implemented using a Calzetti et al. law \citep{calzetti}. The purpose of the varying e-folding times is to account for galaxies with extended star formation histories - an especially important feature for mergers as they often have rich recent star formation.

Figure \ref{plot_kmass} shows the volume-limited mass distributions of galaxies in the merger and control samples. We find that across almost all environments, the spiral-galaxy distributions appear to be virtually the same for both the mergers and the control sample. By contrast, the ellipticals in mergers appear slightly more massive than their control counterparts with a difference in the means of $\sim 2$ dex. This closely parallels the previous conclusion that merger and control spirals occupy similar environments whereas ellipticals in mergers are located in slightly denser environments (which in turn host more massive galaxies on average) than their control counterparts (see \S \ref{enviro}). {\color{black}However, it is important to note that part of this affect could be connected with the different criteria used to distinguish morphologies (the mergers were decided by DWD, D09a, whereas the control morphologies are determined directly from GZ; see \S \ref{morphs}). The effect holds true even when we restrict the merger sample to `sure' morphologies (see the left hand column of Figure \ref{plot_kmass}) and implies that the control-galaxy morphologies allow slightly bluer systems to be classified as ellipticals.

Again though, like with the environment, we find that when we decline to split the merger and control populations by morphology, we do get very similar mass distributions for the merger and control samples while showing slight favour of merging galaxies being more massive (see Appendix \ref{massNenviro}). This should be taken to imply that merging galaxies {\it are} in fact more massive on average than non-merging galaxies, especially since spirals are over-observed in mergers and, being less massive on average compared to ellipticals, should make the average mass of merging galaxies {\it less} than that of the global population (all else being equal). The fact that mergers favour spirals (which are generally less massive) {\it yet} possess an overall distribution just as massive (if not slightly more) than the control sample strongly suggests that galaxies observed in mergers really are more massive.  The more tentative conclusion that this is especially true of ellipticals (by $\sim 2$dex) would corroborate the findings of \citet{bundy2}.}\footnote{Tentative because of the different methods for distinguishing morphologies employed here.}  

\subsection{The Mass-Colour-Morphology Relation}
\begin{figure*}
	\includegraphics[width=150mm]{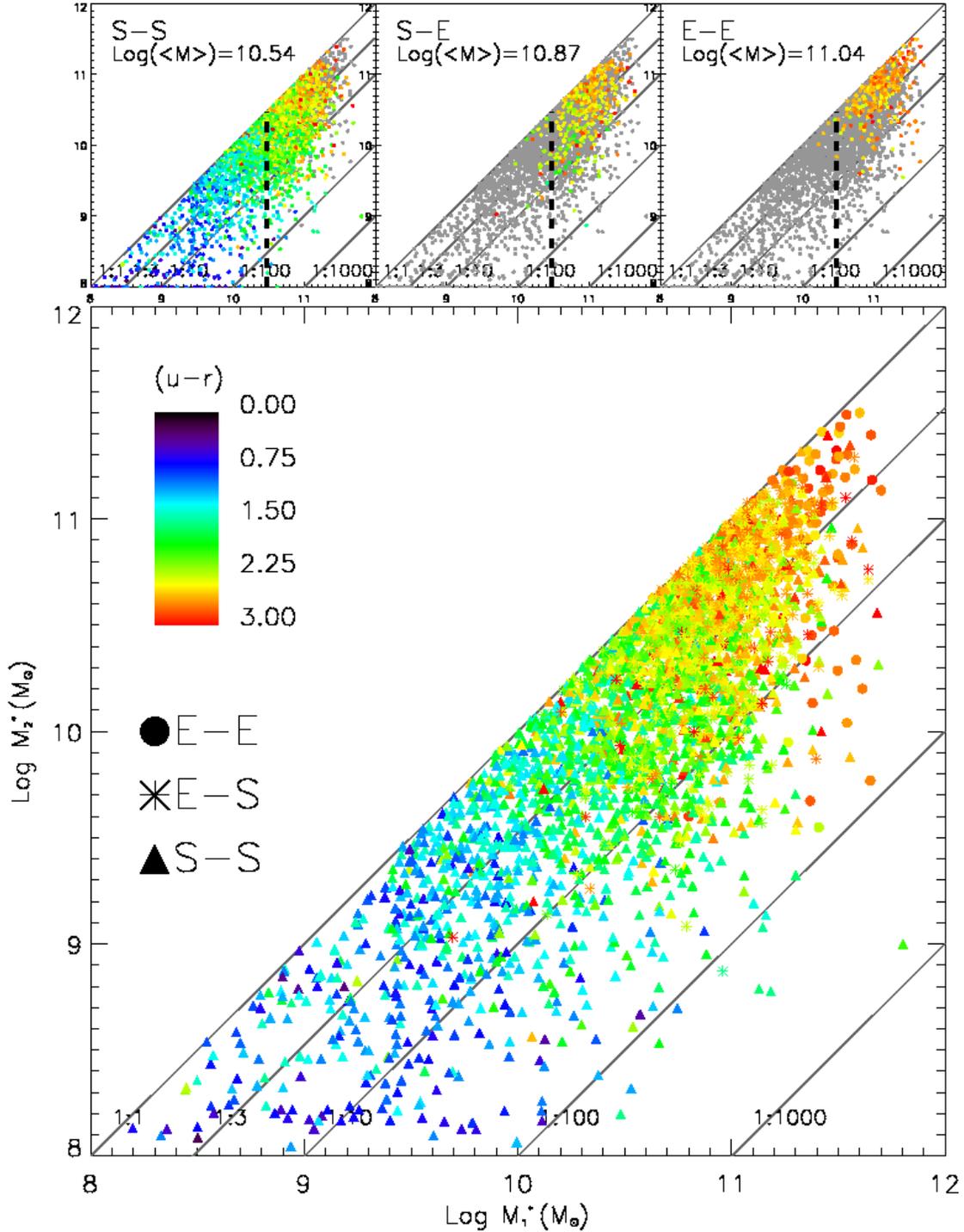}
	\caption[Figure 11]{Mass-Colour-Morphology diagram. {\color{black}Each of the 3003} points represents a merger pair with the more massive galaxy mass plotted on the x-axis and its partner's mass along the y-axis. The colour of each point is the {\it mean} ($u-r$) of the two galaxies. The width of the symbols is the same as the mean-mass error for the entire sample. The symbol represents the morphologies of the galaxies (for S-E we do not distinguish which type is the more massive). We do not impose the magnitude limit on the sample (which would exclude most points for $<10^{10}\mbox{M}_{\astrosun}$) in order to maximise the range of view of the mass-colour-morphology relation. The upper panels individually show the morphological categories over the total merger population (coloured grey). The broken lines therein lie at $3\times10^{10}\mbox{M}_{\astrosun}$ below which ellipticals are rare.}	
	\label{kmass}
\end{figure*}

Figure \ref{kmass} shows the entire merger-pairs catalogue in mass-colour-morphology space. Both colour and morphology scale strongly and smoothly with mass: spiral-spiral mergers dominate the lower-mass end, elliptical-elliptical mergers the upper-mass end and elliptical-spiral mergers roughly in between. A sharp transition mass for galaxy properties within SDSS was noted by \citet{kauffmann1} at $3\times10^{10}\mbox{M}_{\astrosun}$ above which galaxies have ``high surface mass densities, high concentration indices typical of bulges and predominantly old stellar populations'' and below which galaxies have generally opposite characteristics. We find that below this value, ellipticals are extremely rare and above it, spirals are both reddening and diminishing in number in mergers.

The near absence of ellipticals with masses below $3\times10^{10}\mbox{M}_{\astrosun}$ raises the question as to what becomes of the numerous low-mass spiral-spiral mergers we observe. \citet{kauffmann1} noted this special mass with respect to galaxy properties but said little about the mechanism that drives the transition beyond suggesting relations between star formation, feedback mechanisms and halo mass. We hypothesise that this mass could represent a {\it merger} transition related to spiral-spiral survival in major mergers: below this stellar mass, spirals tend to survive mergers, above it they are likely to form an elliptical remnant. Why might this be so? 

The relatively high gas content in low-mass spirals could be the key. The simulations studied in \citet{hopkins} emphasise the role of the progenitor gas-to-stellar-mass ratio as well as feedback mechanisms that serve to retain gas at large radii during the merger process. These outer gas supplies retain angular momentum and aid the reformation of a disc in the post-merger remnant. The transition mass at $3\times10^{10}\mbox{M}_{\astrosun}$ could therefore correspond to some critical gas-to-stellar-mass ratio for disc galaxies. 

On this hypothesis then, galaxies with stellar mass $<3\times10^{10}\mbox{M}_{\astrosun}$ generally have sufficient gas content to bring about disc reformation after a (major-) merger. As spirals increase in stellar mass (at the general expense of gas supply) they become increasingly prone to catastrophic angular-momentum loss with respect to disc maintenance in the event of a merger. Their remaining gas supplies then plunge into the central core and transfer the angular momentum required for disc morphology into the stellar dispersion of the remnant bulge (\citealt{kewley1}). The exhaustion of gas not only limits the system's capacity to retain angular momentum at high radii but also leaves little for passive star formation in the remnant. The resultant bulge-dominated galaxy is thereby destined towards an increasingly red and elliptical galaxy-type (barring further gas accumulation through accretion and gas-rich mergers). 

Since feedback mechanisms are important to gas retention in this model we next examine the AGN and star-formation signatures of our mergers.


\section{AGN, Star-Forming and Quiescent Signatures in Galaxy Mergers}
\label{agn_stuff}
\subsection{Ionisation Processes in Mergers}
\label{agn}
\begin{figure*}
	\includegraphics[width=100mm]{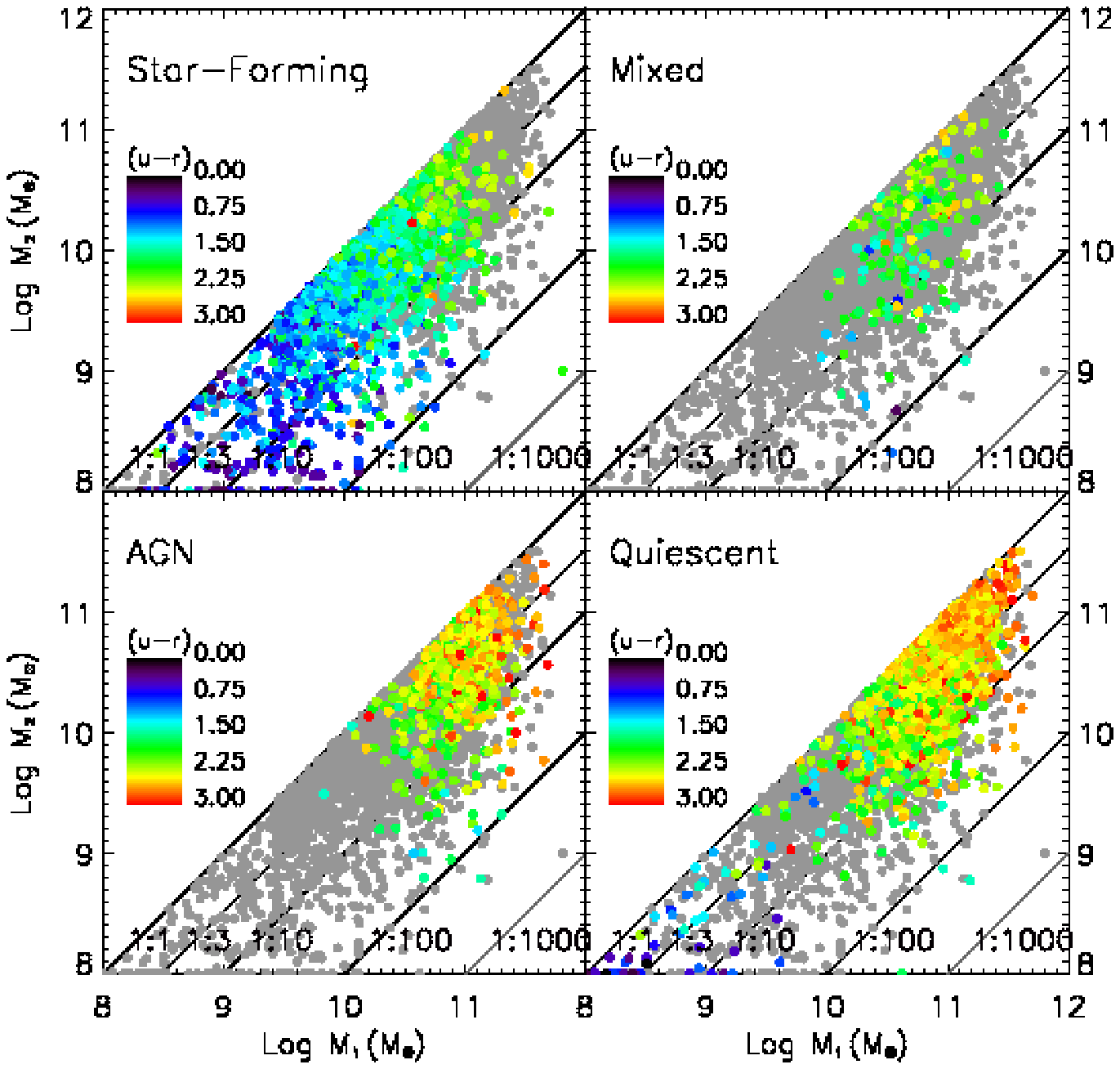}
	\caption[Figure 15]{Spectral Type-Colour-Mass relations. The mass and colour are plotted the same as in figure \ref{kmass} {\color{black}for all 3003 systems}. The panels show the population split into its various spectral types: Quiescent, Star-Forming or AGN. No magnitude limitation is imposed for an enhanced view of mass build-up in relation to these properties.}	
	\label{agn_colour_mass}
\end{figure*}

We perform emission-line diagnostics in order to determine the major sources of ionisation in both the merging and control galaxies. To do this we made use of the publicly available direct fitting tools \texttt{PPXF} and \texttt{GANDALF} (from \citealt{cappellari} and \citealt{sarzi}, respectively) to separate the contribution of the stellar continuum and of the ionised-gas emission to the SDSS spectra, as in \citet{schawinski2}. 

We then used the measured fluxes for the nebular emission lines of our samples to determine the most likely source of ionisation by juxtaposing a number of emission-line ratios as first suggested by \citet{baldwin}. Specifically, we used the reddening-insensitive diagnostic diagrams introduced by \citet{veilleux}, which uses the four optical line ratios [OIII]/H$\beta$, [NII]/H$\alpha$, [SII]/H$\alpha$, and [OI]/H$\alpha$ to separate (i) Star-forming regions, (ii) Seyfert nuclei, (iii) Low-Ionisation Nuclear Emitting regions (LINERs) and (iv) the so-called Mixed/Transition objects, which display the spectral signatures of both HII regions and AGNs. We assigned these classes to all galaxies with S/N$>3$ in at least all the H$\alpha$, H$\beta$, [OIII] and [NII] lines, and further deemed as (v) Quiescent all those galaxies for which such a criterion was not met. {\color{black} In other words a quiescent object is defined as having {\it at least} one weak emission line and so an alternative label is the `weak emission-line' category}. To separate the different kinds of central activity in our merger galaxies we followed the demarcations between purely star-forming systems, transition objects and truly active nuclei drawn by \citet{kauffmann2} and \citet{kewley2}. {\color{black} We combine AGN types into a single category for this presentation}. 

Thus, to each spectral object we assign one of the following classifications:

\begin{enumerate}
\item Star-Forming
\item Mixed (both star-formation and AGN activity)
\item AGN (either Seyfert or LINER)
\item Quiescent
\end{enumerate}

We refer to these possibilities as the galaxy's (ionisation-) `type.' We obtain classifications for 1371 individual galaxies in our volume-limited merger sample. Figure \ref{agn_colour_mass} illustrates the location of these ionisation-types in the same mass-colour space used in Figure \ref{kmass} with no magnitude limitation (a proper volume-limited sample would see decreased numbers of points in the $\mbox{M}\la10^{10}\mbox{M}_{\astrosun}$ regions). Comparing Figures \ref{agn_colour_mass} and \ref{kmass}, we see that the star-forming types occupy the smaller-mass regions which are dominated by spirals and the quiescent types occupy the higher-mass regions, which are dominated by ellipticals. The AGN categories seem to occupy the intermediate-mass regions. 

The lack of star formation and AGN activity in high-mass galaxies suggests that their fuel supply has been exhausted whereas the lack of AGN activity in low-mass gas-rich galaxies suggests that either AGN do not form there (perhaps because they have insufficiently massive black-holes at their centres to generate substantial ionisation) {\it or} that their AGN signatures are obscured by the high gas content and star-formation rates (SFRs) (obscuration of ionisation signatures is a perennial problem of BPT-style classifications; \citealt{baldwin}; \citealt{bamford2}). 


\begin{table*}
\begin{minipage}{126mm}
	\label{agn_table}
	\caption[Table 1]{Percentages of ionisation-types for volume-limited merger and control galaxies. Numbers given are rounded to nearest integer. The `All AGN' row is the sum of the Mixed and AGN percentages. The `All SF' row is the sum of the Mixed and Star-Forming percentages. We include the sample sizes plus Poisson-Counting errors rounded up to the nearest percent.}	
	\begin{tabular}{|| c | c | c | c | c | c | c | c | c ||}
 	\hline                       
	Type 			& S+E Galaxies	& Control	& & S			& E			& & Con.-Spirals	& Con.-Ellipticals 		\\
	\hline 
  	Star-Forming 	& 45$\pm2$		& 14$\pm1$	& & 51$\pm2$	& 0$\pm1$ 	& & 25$\pm2$		& 6	$\pm1$				\\
  	Mixed	 		& 7 $\pm1$		& 4 $\pm1$	& & 8 $\pm1$	& 0$\pm1$	& & 6	$\pm1$		& 2	$\pm1$				\\
  	AGN 			& 16 $\pm1$	& 20$\pm1$	& & 15 $\pm1$	&18$\pm3$	& & 23$\pm2$		& 20$\pm2$				\\
  	Quiescent 		& 32$\pm2$		& 62$\pm2$	& & 26$\pm1$	&81$\pm7$	& & 46$\pm3$		& 73$\pm3$				\\
	\hline
  	All SF		 	& 52$\pm2$		& 18$\pm1$	& & 59$\pm2$	& 0$\pm1$	& & 31$\pm2$		& 8	$\pm1$				\\
  	All AGN		 	& 23$\pm1$		& 24$\pm1$	& & 23$\pm1$	&18$\pm3$	& & 29$\pm2$		& 20$\pm2$				\\
  \hline  
Galaxies in Sample	& 1371			& 1200		& & 1219		& 152		& & 600				& 600					\\
\end{tabular}
\end{minipage}
\end{table*}

{\color{black}
The sample fractions with Poisson counting errors for these various ionisation-types are shown in Table \ref{agn_table}. We exclude `unsure' morphologies (EU, SU). When EU and SU morphologies are included, the percentage of star-forming types decreases by $\sim10\%$ with the quiescent and AGN categories increasing by $\sim5\%$. This effect is to be expected since the `unsure' morphologies include a higher proportion of ellipticals (D09a) which, as the table shows, have fewer star-forming types but more AGN and quiescent. The control sample here consists of 1200 randomly selected volume-limited objects. We also looked at the percentages of the first 600 control galaxies that are deemed to be spirals according to the criteria given in \S\ref{morphs} and likewise for the first 600 control-ellipticals. }

Examining Table \ref{agn_table}, we find that the fraction of AGN in mergers appears no different from the control sample ($\sim23\pm1\%$ compared to $\sim24\pm1\%$) for the total populations. However, splitting the merger and control samples into separate morphologies suggests that the fraction of AGN in merging {\it spirals} is slightly less than in their control counterparts ($\sim23\pm1\%$ compared to $\sim29\pm2\%$). As mentioned though, AGN signatures might be obscured by high star-formation rates (SFRs) and disrupted gas content in merging galaxies. These star-formation rates are seen to be extremely high for merging spirals ($59\pm2\%$) compared to control spirals ($31\pm2\%$). When we further split the merger populations into the three visually-allotted merger `stages' (`separated', `interacting' and `approaching post-merger' see \S \ref{vis_stage}), we find that the percentage of star-forming spiral galaxies in mergers for these stages are $59\pm8\%$, $50\pm2\%$ and $43\pm7\%$ respectively. The descending percentages suggest that star-formation takes place early-on in the merger process. When we examine the fractions of AGN types in spirals for these stages we obtain $21\pm5\%$, $23\pm2\%$ and $32\pm6\%$ which shows slight signs of ascending AGN activity within merging spirals as they approach the post-merger stage. Alternatively, this could suggest that where SFRs are less intense, AGN signatures become easier to detect or even that we are seeing the effects of AGN feedback quenching SF. 

The sample of ellipticals in mergers, by contrast, resembles that of the control ellipticals when split into ionisation types except that {\it none} of them are star-forming types. Both merging and control ellipticals are dominated by quiescent types ($81\pm7\%$ and $73\pm3\%$) and have the same fraction of AGN ($18\pm3\%$ and $20\pm2\%$). In short, the internal properties of ellipticals appear basically unaffected by the merger process and thus live up to the `red-and-dead' stereotype, dominating the quiescent category. 
 
Our results are in broad agreement with previous studies. Induced star formation in interacting galaxies was first quantified by \citet{keel} and \citet{kennicutt1}. They claimed, however, that both star formation and nuclear activity is enhanced in close-pairs. Similar studies since then have strongly confirmed that mergers-induce star formation (see \S \ref{intro}; \citealt{hopkinsAM}), though the idea that AGN are significantly induced by mergers remains tentative (\citealt{ellison}). 

Our study so far strongly confirms that mergers significantly enhance SFRs but {\it only} in spirals - there appears to be no effect at all upon our visually-inspected `sure' ellipticals. Our work also lends very little support to the notion that AGN activity is enhanced by the merger process, the one exception perhaps being in late-stage spirals.

\subsection{Star-Forming Rates in Merging Galaxies}
\label{sfr}
\begin{figure}
	\includegraphics[width=84mm]{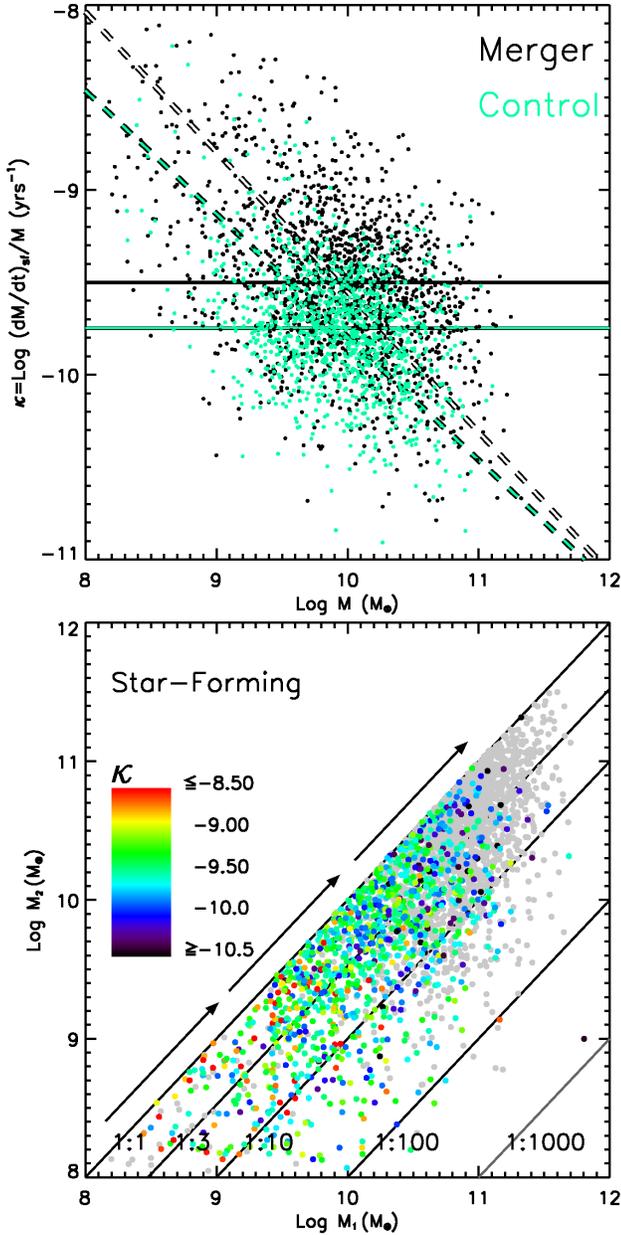}
	\caption[Figure 16]{The specific-star-formation rate ($\kappa$) compared to stellar-mass. The upper panel shows the relationship between stellar mass and specific-star-formation ($\kappa$) for the {\color{black}1588 star-forming galaxies in mergers in our catalogue and the same number of star-forming systems taken from the control sample}. The broken lines are linear best-fits to the samples. The solid horizontal lines show $<\kappa>$ for the two populations. The lower panel shows $\kappa$-mass space for the star-forming merger systems (any system with at least one galaxy of the star-forming type) where the colour scale represents $\kappa$ of the galaxy which is star-forming. Neither sample is volume-limited for an enhanced general overview of $\kappa$ across our stellar-mass range. The arrows indicate the gas-depletion evolution advocated in this study.}	
	\label{sfr_kmass}
\end{figure}

Having found the fraction of galaxies classified as `star-forming,' we now wish to quantify the {\it rates} at which their star formation occurs. We use the integrated spectral flux of the extinction-corrected $\mbox{H}\alpha$ lines derived by our ionisation-types assessment to obtain an absolute rest-frame flux. We scale the flux measured in the $3''$-diameter SDSS fibre aperture to give an estimate of the total flux, using the ratio of Petrosian and $3''$ aperture fluxes in the $r$-band photometry. We then apply the model $\mbox{H}\alpha$-SFR relation derived by \citet{kennicutt2}, Eq. 3:

$$ SFR_{\mbox{H}\alpha}\, [\mbox{M}_{\astrosun}yrs^{-1}] = 7.94\times10^{-42}L_{\mbox{H}\alpha} \,[\mbox{ergs/s}]$$

to obtain estimates for the SFRs of our merging and control galaxies. We find that our (volume-limited) sample of star-forming merger galaxies has a mean SFR$_{\mbox{H}\alpha}$ of $\sim5.2\mbox{M}_{\astrosun} yr^{-1}$. The equivalent control sample has $\sim2.6\mbox{M}_{\astrosun} yr^{-1}$, i.e. the merger process enhances our SFRs by a factor $\sim 2$. {\color{black}The highest SFR for any of our star-forming merging galaxies is $\sim95\mbox{M}_{\astrosun} yr^{-1}$.}

However, our sample involves a range in masses over $3-4$ orders of magnitude and so it is not entirely appropriate to compare SFRs across such a range (one would expect larger galaxies to have a greater absolute SFR) and so we quantify the {\it relative} size of the SFR for each galaxy by defining the {\it specific-}SFR, $\kappa$, as the log of the SFR$_{\mbox{H}\alpha}$ per stellar mass unit:

\begin{equation}
\label{kappa}
\kappa  = \log \left[ \frac{(\frac{\mbox{dM}}{\mbox{dt}})_{star formation}}{\mbox{M}^*}\right]   \mbox{yr}^{-1}.
\end{equation}

The upper panel of Figure \ref{sfr_kmass} plots the value of $\kappa$ against the stellar mass for each galaxy of the star-forming type. There is a negative correlation between these two quantities such that the more massive a star-forming galaxy is, the smaller its SFR$_{\mbox{H}\alpha}$ per stellar mass (similar to the findings of \citealt{brinchmann} Figure 13). The star-forming control sample has a mean $\kappa$ of $\sim0.25$ less than the star-forming mergers. The control's $\kappa-\mbox{M}^*$ gradient is also shallower indicating that as the size of the galaxy increases, the relative star formation enhancement induced by the merger diminishes. This is not surprising since gas supply in galaxies should generally scale down with stellar mass (\citealt{noeske}). Taking $\kappa$ as a proxy for gas content (reminiscent to the Schmidt law \citet{schmidt}), we can interpret this relation to mean that the larger the star-forming galaxy becomes (with respect to stellar mass) so their merging becomes `drier.' By the time the gas is completely exhausted, there simply is no fuel available for SF, even in a merger. 

These observations therefore lend well to the hypothesis that a critical gas-to-stellar-mass ratio exists for spirals which could correspond to galaxies with the $3\times10^{10}\mbox{M}_{\astrosun}$ mass of \citet{kauffmann1}. By taking $\kappa$ as a proxy for relative gas abundance, one can envisage a smoothly decreasing gas supply with increasing stellar mass in the lower panel of Figure \ref{sfr_kmass}. The number of star-forming spirals then begins to diminish for those of mass beyond $3\times10^{10}\mbox{M}_{\astrosun}$ which is roughly where elliptical galaxies begin to take over (Figure \ref{kmass}). The suggestion is that spirals with gas supply corresponding to {\color{black} $\kappa\la10^{-10}$} will most likely result in an elliptical remnant should they undergo a major merger, the likelihood of this result scaling up with decreasing $\kappa$. 


\section{Summary and Discussion}
\label{conc}

In the previous study, D09a, we found that the spiral-to-elliptical ratio in mergers ($N_s/N_e$) was high by a factor of at least 2 in our sample compared to the global population. The first aim of this paper was to discern the likely cause for this discrepancy suggesting that it is either the result of an environmental preference for mergers to take place where spirals are relatively abundant {\it or} that the time-scales of detectability for spiral-mergers are longer than those for elliptical-mergers.


To test the role of environment we used the adaptive-Gaussian-environment parameter $\rho_{\sigma}$ to create distributions for our samples. By comparison to the randomly selected control sample of SDSS galaxies with spectra, we found that mergers occupy similar if not {\it denser} environments than the control sample. This, if anything, would not favour the presence of spirals in mergers but ellipticals since it is known that denser environments are favourable to elliptical galaxies (\citealt{dressler}). 

We concluded therefore that the high number of spirals in mergers is unlikely to be an environmental effect. On the other hand, the suggested alternative (that the time scales for a merger to reach a relaxed state vary depending on the internal properties of the galaxies) seems intrinsically plausible and has been corroborated by other studies (\citealt{bell}; \citealt{lotz2}; \citealt{lotz3}). Spiral galaxies are typified by relatively large gas reservoirs, a more uniform distribution of matter along their radius and lower total mass in comparison to ellipticals. One would expect therefore that, when two ellipticals merge, they tend to produce comparatively faint tidal tails and little star formation, making their detection a more difficult observational task. The role of mass remains unclear though. The simulations of \citet{lotz3} suggested that mass made little difference to these time-scales but our results suggested that very massive ellipticals in mergers were more likely to merge (\S \ref{mass}).  

This slight excess in mass is complimentary to a slight excess in environmental density. Suppose that the probability of a galaxy merging at some general time, $p_m$, is only a function of galaxy mass and environment, $p_m=p_m(\mbox{M}^*,\rho)$.\footnote{This takes into account the number density and peculiar velocities of surrounding galaxies since these are both functions (or definitions) of environmental measure.} For any given environment, $\rho$, a more massive galaxy exerts a stronger pull on its neighbours and so, all else being equal, a more massive galaxy should be more likely to merge.\footnote{Moreover, for any environment taken as a closed system orbiting a common centre of mass where it can be assumed that it's constituent bodies are in equilibrium (having the same kinetic energy), a more massive body will have a smaller peculiar velocity with respect to the system's centre of mass making it more conducive to gravitational binding with some other orbiting body.} 

More massive galaxies are also more likely to occupy denser environments (given the morphology-environment relationship, e.g. \citet{dressler} and the mass-morphology relationship, e.g. \citet{kauffmann1}) so that two galaxies of mass $\mbox{M}^*_1$ and $\mbox{M}^*_2$ where $\mbox{M}^*_1>\mbox{M}^*_2$ which have the same probability of merging, $p_1(\mbox{M}^*_1,\rho_1)=p_2(\mbox{M}^*_2,\rho_2)$ must occupy different environments and, since mass generally scales with environment, we must have $\rho_1>\rho_2$. In other words, {\it both} the mass {\it and} environment distributions of galaxies in mergers should appear rightward-shifted compared to the global population as we see in Figures \ref{plotEnviro} and \ref{plot_kmass}.

Since environmental factors do not provide an explanation for the high $N_s/N_e$ observed in mergers, we conclude that mergers involving spiral galaxies remain detectable for longer periods. Whereas the study by \citet{lotz3} provided theoretical evidence that this is in fact the case, this study provides empirical evidence that these time-scales of detectability do indeed vary. This should be taken into consideration by those that aim to convert an observed merger fraction to an absolute merger rate for implementation in hierarchical models. 

Mergers with spirals must remain detectable for longer due to their internal properties and so we turned to investigate them, beginning with the photometric properties of mergers. We showed that the colours of merging galaxies scale strongly with mass and morphology and are more spread compared to ordinary galaxies (\S \ref{masses}). In particular, mergers exhibit a strong blue tail which we concluded is due to intense star formation induced by the merger process. 

Below the stellar transition mass $\sim3\times10^{10}\mbox{M}_{\astrosun}$ noted by \citet{kauffmann1} we found that ellipticals were rare in both the merger and the control samples though spirals were fairly common in both. What then becomes of the numerous low-mass spiral-spiral mergers? It was posited in \S \ref{intro} that at least some spiral-spiral mergers survive major mergers and this lead to the hypothesise that the transition mass of $\sim3\times10^{10}\mbox{M}_{\astrosun}$ corresponds to a transition between general-disc survival and general-disc destruction in mergers. 

Such a transition would be closely linked with gas dynamics in mergers. Simulations studying disc survival have placed great emphasis on the interactions between gas and stars in mergers suggesting that galaxies with high gas-to-stellar-mass ratios and reservoirs at high radii are highly capable of rapid disc-reformation after dynamical relaxation (\citealt{hopkins}). As spiral galaxies evolve they expend gas in their disc via passive star formation and merger-induced drainage leading to an increasingly lower gas-to-stellar-mass ratio. Since the gas content in spirals generally scales down with stellar mass, there must be some average gas-to-stellar-mass ratio for spirals at $3\times10^{10}\mbox{M}_{\astrosun}$ and this, we hypothesise, marks a critical point beyond which spirals are unlikely to survive major mergers.

While the gas-to-stellar-mass ratio is important, it cannot be the sole determinant of disc survival. For example the distribution of gas is also an important factor meaning that feedback mechanisms that retain gas at high radii are indirectly involved in disc survival/destruction in mergers. This prompted AGN-SFR analysis using the spectral-line widths available to our catalogue. 

We found that mergers induce intense star formation but {\it only} in mergers involving spirals (see Table \ref{agn_table}) - ellipticals are hardly affected and dominate the quiescent category. This fits with the `red-and-dead' stereotype for giant ellipticals and suggests that mergers can account well for the spread towards the high-mass end of galaxies in the red-colour sequence (i.e. giant elliptical-elliptical mergers increase the luminosity of the progenitor, but negligibly affect its colour and internal properties). 

By contrast, we found little overall evidence for increased AGN activity in mergers in broad agreement with several recent studies (\citealt{barton2}; \citealt{alonso}; \citealt{li}) though contrary to early reports such as \citet{kennicutt1} and, more recently, \citet{woods}, \citet{schawinski6}. The recent study by \citet{ellison} also found little evidence for increased AGN activity in their close-pairs sample and concluded that, if AGN are induced by mergers, then they must occur at stages later than close-pairs typically examine. In D09a, we did show in fact that our merger-location technique picks up mergers in later stages compared to the close-pairs technique. Furthermore, when we divided our mergers into their visually assigned stages, there appeared to be a slight increase in the proportion of merging galaxies in the `approaching post-merger' stage ($32\pm6\%$ with mean projected core-separation $\sim5$kpc) in comparison to mergers at earlier stages ($23\pm2\%$ with mean projected core-separation $\sim13$kpc). Caution is urged here though since the counting errors are large and there may be obscuration affects associated with strong star formation signatures.

We found that the specific-SFRs (defined in Eq. (\ref{kappa}), \S \ref{sfr}) are higher in star-forming mergers, on average, than in the star-forming control galaxies by $\sim2$. For the star-forming galaxies in mergers we find that the specific-star-formation rate scales down with stellar mass. We interpreted this to mean that gas supply is being continually drained as galaxies accumulate stellar-mass. This is consistent with the hypothesis that a critical gas-to-stellar-mass ratio emerges near $3\times10^{10}\mbox{M}_{\astrosun}$ for disc survival/destruction.

The results of this study generally imply that, where mergers do happen, their effects are powerful on spirals (eroding their gas and angular momentum supplies and strongly enhancing their SFRs) but much weaker on ellipticals. This in turn affects the time-scales of detectability for mergers which should be taken into account by studies aiming to convert merger fractions into merger rates. 

Many interesting clues about galaxy evolution can be gleaned from our data and future projects such as Galaxy Zoo Two applied to SDSS and higher redshift surveys promise exciting results.   

\section{Acknowledgments}
\label{ack}

D. W. Darg acknowledges funding from the John Templeton Foundation. S. Kaviraj acknowledges a Research Fellowship from the Royal Commission for the Exhibition of 1851 (from Oct 2008), a Leverhulme Early-Career Fellowship (till Oct 2008), a Senior Research Fellowship from Worcester College, Oxford and support from the BIPAC institute, Oxford. C. J. Lintott acknowledges funding from The Levehulme Trust and the STFC Science in Society Program. K. Schawinski was supported by the Henry Skynner Junior Research Fellowship at Balliol College, Oxford.

Funding for the SDSS and SDSS-II has been provided
by the Alfred P. Sloan Foundation, the Participating Institutions,
the National Science Foundation, the U.S. Department
of Energy, the National Aeronautics and Space
Administration, the Japanese Monbukagakusho, the Max
Planck Society, and the Higher Education Funding Council
for England. The SDSS Web Site is http://www.sdss.org/.
The SDSS is managed by the Astrophysical Research Consortium for the Participating Institutions. The Participating
Institutions are the American Museum of Natural
History, Astrophysical Institute Potsdam, University of
Basel, University of Cambridge, Case Western Reserve University,
University of Chicago, Drexel University, Fermilab,
the Institute for Advanced Study, the Japan Participation
Group, Johns Hopkins University, the Joint Institute for
Nuclear Astrophysics, the Kavli Institute for Particle Astrophysics
and Cosmology, the Korean Scientist Group, the
Chinese Academy of Sciences (LAMOST), Los Alamos National
Laboratory, the Max-Planck-Institute for Astronomy
(MPIA), the Max-Planck-Institute for Astrophysics (MPA),
New Mexico State University, Ohio State University, University
of Pittsburgh, University of Portsmouth, Princeton
University, the United States Naval Observatory, and the
University of Washington.


\appendix

\section{Environment and Masses of Combined Morphologies for Merger and Control Sample}
\label{massNenviro}

It was claimed in \S \ref{enviro} with reference to Figure \ref{plotEnviro} that merging galaxies appeared to occupy very similar, if not, denser environments for both ellipticals and spirals compared to their control counterparts. A very similar set of results was claimed in \S \ref{mass} with reference to Figure \ref{plot_kmass} suggesting that merging galaxies possessed very similar, if not, more massive stellar masses than their control counterparts. In both cases, the excesses appeared slightly higher for ellipticals, though still only with a $\sim2$dex difference in the means for both $\rho_{g}$ and M$^{*}$.

\begin{figure}
	\includegraphics[width=70mm]{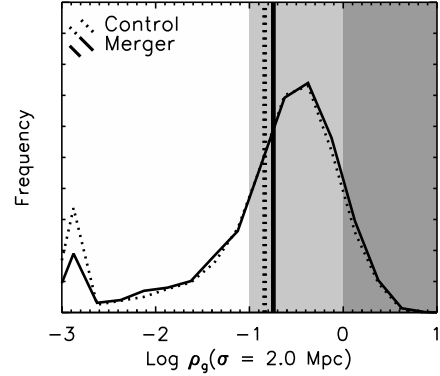}
	\caption[Figure App2]{Combined-morphologies version of Figure \ref{plotEnviro}.}	
	\label{plot_enviro_combine}
\end{figure}

\begin{figure}
	\includegraphics[width=70mm]{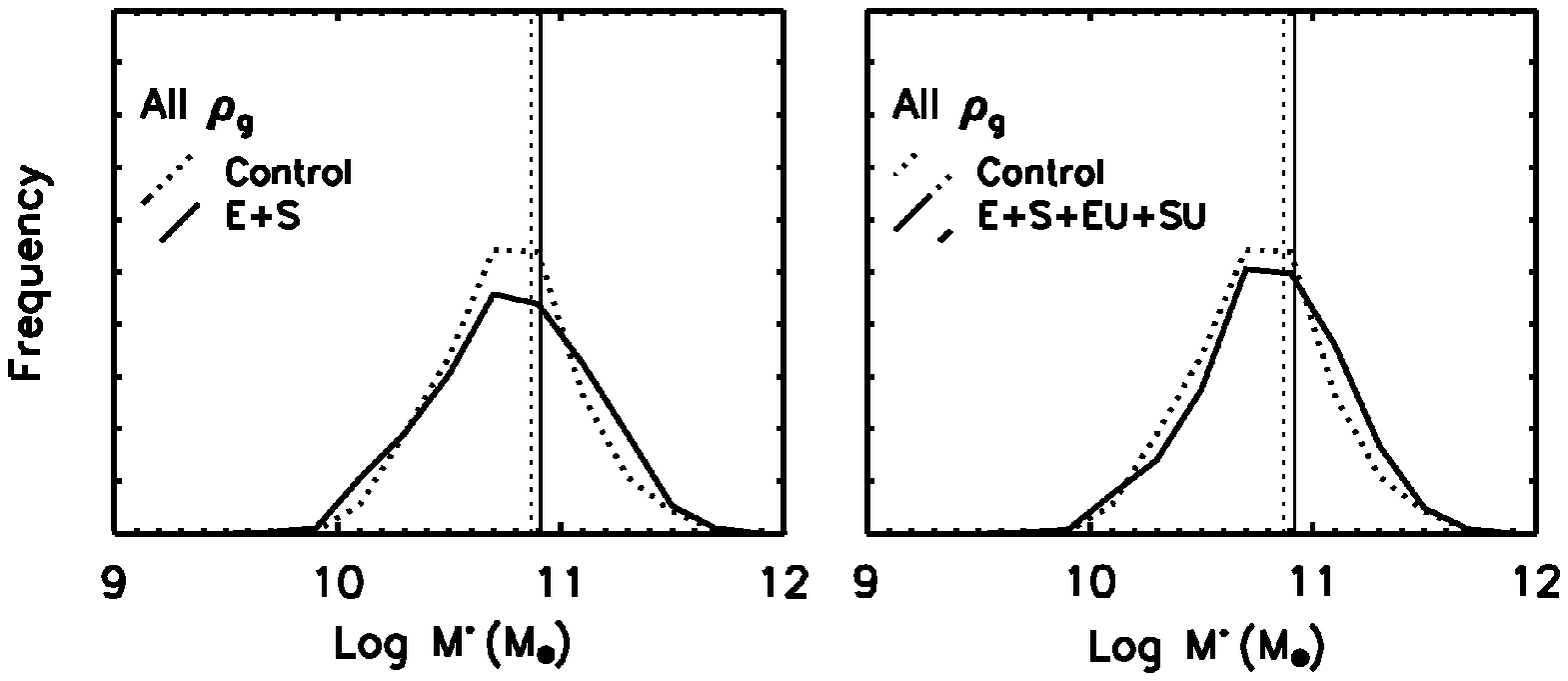}
	\caption[Figure App3]{Combined-morphologies version of Figure \ref{plot_kmass}.}	
	\label{plot_kmass_combine}
\end{figure}

However, the morphologies for the two samples were selected by different means (the mergers visually by DWD and the control sample by GZ data as described in \S \ref{morphs}). In particular the control sample, when divided into ellipticals and spirals, was rendered incomplete by the stipulation that $\left|f_e-f_s \right|<0.1$. We therefore reproduce Figures \ref{plotEnviro} and \ref{plot_kmass} in Figures \ref{plot_enviro_combine} and \ref{plot_kmass_combine} without distinguishing morphologies.

These figures confirm the basic result that for both environment and stellar masses, the merger and control distributions are very similar with the mergers exhibiting a very slight excess in both cases with respect to their mean values. However this is significant since, as argued in D09a, the merger sample has a high spiral-to-elliptical ratio compared to the global population which is represented here by the control sample. This should decrease the mean-values of the mergers for both environment and stellar mass since spirals are known to be less massive and occupy less dense environments than ellipticals. To put this in other words, the mergers manage to `keep up' with the control sample despite the handicap of a high spiral population which strengthens the claim that mergers do in fact have a slight excess in mass and environmental density compared with the global population.



\begin{thebibliography}{99}


\bibitem[\protect\citeauthoryear{Alonso et al.}
{2007}]{alonso} Alonso M. S., Lambas D. G., Tissera P. B., \& Coldwell G. 2007, MNRAS, 375, 1017.

\bibitem[\protect\citeauthoryear{Baldwin, Phillips \& Terlevich}
{1981}]{baldwin} Baldwin, J. A., Phillips M. M., Terlevich R., 1981, PASP, 93, 5B.



\bibitem[\protect\citeauthoryear{Bamford et al.}
{2008}]{bamford2} Bamford S. P., Rojas A. L., Nichol R. C., Miller C. J., Wasserman L., Genovese C. R., Freeman P. E., 2008, MNRAS, 391, 607

\bibitem[\protect\citeauthoryear{Bamford et al.}
{2009}]{bamford} Bamford S. P. et al., 2009, MNRAS, 393, 1324

\bibitem[\protect\citeauthoryear{Barnes \& Hernquist}
{1996}]{barnes} Barnes J. E. \&  Hernquist L., 1996, ApJ, 471, 115.

\bibitem[\protect\citeauthoryear{Barton, Geller \& Kenyon}
{2000}]{barton2} Barton E. J., Geller M. J., \& Kenyon S. J. 2000, ApJ, 530, 660.


\bibitem[\protect\citeauthoryear{Bell et al.}
{2006}]{bell} Bell E. F., Phleps S., Somerville R. S., Wolf C., Borch A., Meisenheimer K., 2006, ApJ, 652, 270B.

\bibitem[\protect\citeauthoryear{Benson et al.}
{2002}]{benson1} Benson A. J., Lacey C. G., Baugh C. M., Cole S., Frenk C. S., 2002, MNRAS, 333, 156B.

\bibitem[\protect\citeauthoryear{Benson et al.}
{2003}]{benson2} Benson A. J., Bower R. G., Frenk C. S., Lacey C. G., Baugh C. M., Cole S., 2003, ApJ, 599, 38B.

\bibitem[\protect\citeauthoryear{Bernardi et al.}
{2003}]{bernardi} Bernardi M., et al., 2003, AJ, 125, 1817B.


\bibitem[\protect\citeauthoryear{Blanton et al.}
{2003}]{blanton} Blanton M. R., et al., 2003, AJ, 125, 2348B.

\bibitem[\protect\citeauthoryear{Blumenthal et al.}
{1984}]{blum} Blumenthal G. R., Faber S. M., Primack J. R., Rees M. J., 1984, Nature, 311, 517B. 


\bibitem[\protect\citeauthoryear{Brinchmann et al.}
{2004}]{brinchmann} Brinchmann J., Charlot S., White S. D. M., Tremonti C., Kauffmann G., Heckman T., Brinkmann J., 2004, MNRAS, 351, 1151.  


\bibitem[\protect\citeauthoryear{Bundy et al.}
{2009}]{bundy2} Bundy K., Fukugita M., Ellis R. S., Targett T. A., Belli S., Kodama T., 2009, ApJ, 697, 1369 

\bibitem[\protect\citeauthoryear{Calzetti}
{2000}]{calzetti} Calzetti D., Armus L., Bohlin R. C., Kinney A. L., Koornneef J., Storchi-Bergmann T., ApJ, 533, 682


\bibitem[\protect\citeauthoryear{Cappellari \& Emsellem}
{2004}]{cappellari} Cappellari M. \& Emsellem E., 2004, PASP, 116, 138


\bibitem[\protect\citeauthoryear{Christlein \& Zabludoff}
{2005}]{christlein} Christlein D. \& Zabludoff A. I., 2005, ApJ, 621, 201


\bibitem[\protect\citeauthoryear{Conselice}
{2003}]{con2003} Conselice C. J., 2003, ApJS, 147, 1 




\bibitem[\protect\citeauthoryear{Conselice, Rajgor \& Myers}
{2008}]{con2008} Conselice C. J., Rajgor S., Myers R., 2008, MNRAS, 386, 909


\bibitem[\protect\citeauthoryear{Daddi et al.}
{2004}]{dad} Daddi E. et al., 2004, ApJ, 600, L127   

\bibitem[\protect\citeauthoryear{Darg et al.}
{2009}]{darg} Darg D. W. et al., 2009, MNRAS, submitted (D09a in paper)   


\bibitem[\protect\citeauthoryear{De Propos et al.}
{2008}]{depropris} De Propris R., Conselice C. J., Liske J., Driver S. P., Patton D. R., Graham A. W., Allen P. D., 2008, ApJ, 666, 212


\bibitem[\protect\citeauthoryear{Dressler}
{1980}]{dressler} Dressler A., 1980, AJ, 236, 351   

\bibitem[\protect\citeauthoryear{Ellis \& Silk}
{2007}]{ellis} Ellis R. \& Silk J., 2007, arXiv:0712.2865

\bibitem[\protect\citeauthoryear{Ellison et al.}
{2008}]{ellison} Ellison S. L., Patton D. R., Simard L., McConnachie A. W., 2008, AJ, 135, 1877

\bibitem[\protect\citeauthoryear{Eggen et al.}
{1962}]{eggen} Eggen O. J. et al., 1962, ApJ, 136, 748   



\bibitem[\protect\citeauthoryear{Gebhardt et al.}
{2000}]{gebhardt} Gebhardt et al., ApJ, 2000, 539, 13

\bibitem[\protect\citeauthoryear{Genzel et al.}
{2008}]{genzel} Genzel et al., ApJ, 2008, 563, 527   


\bibitem[\protect\citeauthoryear{Hopkins et al.}
{2003}]{hopkinsAM} Hopkins A. M. et al., 2003, ApJ, 599, 971

\bibitem[\protect\citeauthoryear{Hopkins et al.}
{2009a}]{hopkins} Hopkins P. F., Cox T. J., Younger J. D., Hernquist L., 2009a, ApJ, 691, 1168
 
\bibitem[\protect\citeauthoryear{Hopkins et al.}
{2009b}]{hopkins2} Hopkins P. F., Somerville R. S., Cox T. J., Hernquist L., Jogee S, Kereš D., Ma C.-P., Robertson B., Stewart K., 2009b, MNRAS, 397, 802H



\bibitem[\protect\citeauthoryear{Jogee}
{2008}]{jog2} Jogee S., 2008, arXiv:0408383


\bibitem[\protect\citeauthoryear{Kaviraj}
{2007a}]{kaviraj} Kaviraj S., 2007, arXiv:0801.0127v1

\bibitem[\protect\citeauthoryear{Kaviraj}
{2007b}]{kaviraj2} Kaviraj S., 2007, arXiv:0710.1311v1

\bibitem[\protect\citeauthoryear{Kaviraj et al.}
{2007}]{kaviraj3} Kaviraj S., Kirkby L. A., Silk J., Sarzi M., 2007, MNRAS, 382, 960.

\bibitem[\protect\citeauthoryear{Kauffmann et al.}
{2003a}]{kauffmann1} Kauffmann G. et al., 2003a, MNRAS, 341, 54

\bibitem[\protect\citeauthoryear{Kauffmann et al.}
{2003b}]{kauffmann2} Kauffmann G. et al., 2003b, MNRAS, 346, 1055

\bibitem[\protect\citeauthoryear{Keel et al.}
{1985}]{keel} Keel W. C., Kennicutt R. C. Jr., Hummel E., van der Hulst J. M., 1985, AJ, 90, 708  

\bibitem[\protect\citeauthoryear{Kennicutt et al.}
{1987}]{kennicutt1} Kennicutt R. C. Jr., Roettiger K. A., Keel W. C., van der Hulst J. M., Hummel E., 1987, AJ, 93, 1011  

\bibitem[\protect\citeauthoryear{Kennicutt}
{1998}]{kennicutt2} Kennicutt Jr. R. C., 1998, ApJ, 498, 541  

\bibitem[\protect\citeauthoryear{Kewley, Geller \& Barton}
{2006}]{kewley1} Kewley L. J., Geller M. J. \& Barton E. J., 2006, AJ, 131, 2004

\bibitem[\protect\citeauthoryear{Kewley et al.}
{2006}]{kewley2} Kewley L. J., Groves B., Kauffmann G., Heckman T., 2006, MNRAS, 372, 961

\bibitem[\protect\citeauthoryear{Khalatyan et al.}
{2008}]{khalatyan} Khalatyan A., 2008, MNRAS, 387, 13






\bibitem[\protect\citeauthoryear{Larson \& Tinsley}
{1978}]{larson} Larson R. B. \& Tinsley B. M., 1978, ApJ, 219, 46


\bibitem[\protect\citeauthoryear{Li et al.}
{2008}]{li} Li C., Kauffmann G., Heckman T. M., White S. D. M., Jing Y. P., 2008, MNRAS, 385, 1915


\bibitem[\protect\citeauthoryear{Lintott et al.}
{2006}]{lintott2} Lintott C. J. et al., 2006, ApJ, 648, 826

\bibitem[\protect\citeauthoryear{Lintott et al.}
{2008}]{lintott1} Lintott C. J. et al., 2008, 389, 1179L.



\bibitem[\protect\citeauthoryear{Lotz et al.}
{2008a}]{lotz2} Lotz J. M. et al., 2008, ApJ, 672, 177 

\bibitem[\protect\citeauthoryear{Lotz et al.}
{2008b}]{lotz3} Lotz J. M., Jonsson P., Cox T. J., Primack J. R., 2008, MNRAS, 391, 1137L

\bibitem[\protect\citeauthoryear{Nagamine et al.}
{2005}]{nag} Nagamine K., Cen R., Hernquist L., Ostriker J. P., Springel V., 2005, ApJ, 618, 23  


\bibitem[\protect\citeauthoryear{Maraston}
{1998}]{maraston1} Maraston C., 2005, MNRAS, 300, 872

\bibitem[\protect\citeauthoryear{Maraston}
{2005}]{maraston2} Maraston C., 2005, MNRAS, 362, 799

\bibitem[\protect\citeauthoryear{Sanders \& Mirabel}
{1996}]{sanders} 	Sanders D. B. \& Mirabel I. F., 1996, ARA\&A, 34, 749.

\bibitem[\protect\citeauthoryear{Noeske et al.}
{2007}]{noeske} Noeske K. G. et al., 2007, AJ, 660, L47







\bibitem[\protect\citeauthoryear{Sarzi et al.}
{2006}]{sarzi} Sarzi M. et al., 2006, MNRAS,  366, 1151 


\bibitem[\protect\citeauthoryear{Schawinski et al.}
{2006}]{schawinski1} Schawinski K. et al., 2006, Nature, 442, Issue: 7105, 888 

\bibitem[\protect\citeauthoryear{Schawinski et al.}
{2007a}]{schawinski2} Schawinski K., Thomas D., Sarzi M., Maraston C., Kaviraj S., Joo S.-J., Yi S. K., Silk J., 2007a, MNRAS, 382, 1415 

\bibitem[\protect\citeauthoryear{Schawinski et al.}
{2007b}]{schawinski3} Schawinski K. et al., 2007b, ApJS, 173, 512


\bibitem[\protect\citeauthoryear{Schawinski et al.}
{2009a}]{schawinski5} Schawinski K. et al., 2009, ApJ, 690, 1672S

\bibitem[\protect\citeauthoryear{Schawinski et al.}
{2009b}]{schawinski6} Schawinski K., Virani S., Simmons B., Urry C. M., Treister E., Kaviraj S., Kushkuley B., 2009, ApJL, 692, 19S

\bibitem[\protect\citeauthoryear{Schweizer et al.}
{1996}]{schweizer1} Schweizer F., Miller B. W., Whitmore B. C., Fall S. M., 1996, AJ,  112, 1839 

\bibitem[\protect\citeauthoryear{Schweizer et al.}
{2005}]{schweizer2} Schweizer F., 2005, ASSL, 329, 143S

\bibitem[\protect\citeauthoryear{Schweizer et al.}
{2006}]{schweizer3} Schweizer F., 2006, arXiv:0606036

\bibitem[\protect\citeauthoryear{Schmidt}
{1959}]{schmidt} Schmidt M., 1959, ApJ, 129, 243



\bibitem[\protect\citeauthoryear{Somerville}
{2006}]{somerville} Somerville R. S., 2006, AAS,  209, 6201 

\bibitem[\protect\citeauthoryear{Springel et al.}
{2005}]{springel} Springel V. et al., 2005, Nature,  435, 629 

\bibitem[\protect\citeauthoryear{Storchi-Bergmann et al.}
{2001}]{storchi} Storchi-Bergmann T., González Delgado R. M., Schmitt H. R., Cid F. R., Heckman T., 2001, ApJ, 559, 147


\bibitem[\protect\citeauthoryear{Strauss et al.}
{2002}]{strauss} Strauss M. A. et al., 2002, AJ, 124, 1810





\bibitem[\protect\citeauthoryear{Veilleux \& Osterbrock}
{1987}]{veilleux} Veilleux S. \& Osterbrock D. E., 1987, AJSS, 63, 295 


\bibitem[\protect\citeauthoryear{Woods, Geller \& Barton}
{2006}]{woods2} Woods D. F., Geller M. J. \& Barton E. J., 2006, AJ,  132, 197 

\bibitem[\protect\citeauthoryear{Woods \& Geller}
{2007}]{woods} Woods D. F. \& Geller M. J., 2007, AJ,  134, 527 

\bibitem[\protect\citeauthoryear{Worthey, Faber \& Gonzalez}
{1992}]{wor} Worthey G., Faber S. M., Gonzalez J. J., 1992, ApJ,  398, 69 

\bibitem[\protect\citeauthoryear{Yoon et al.}
{2008}]{yoon} Yoon J. H. et al., 2008, ApJS, submitted


\bibitem[\protect\citeauthoryear{Zepf \& Ashman}
{1999}]{zepf} Zepf S. E., Ashman K. M., English J., Freeman K. C., Sharples R. M., 1999, AJ,  118, 752 

\end{thebibliography}
\end{document}